\documentstyle[aps,epsfig]{revtex}

\newcommand{\NL}{\mbox{$N_1$}}

\newcommand{\phiR}{\mbox{$\phi_0$}}
\newcommand{\phiL}{\mbox{$\phi_1$}}

\newcommand{\fL}{\mbox{$f_1$}}

\newcommand{\cL}{\mbox{$c_1$}}

\begin{document}
\sloppy

\title{Dynamical Monte Carlo Study of Equilibrium Polymers (II):\\ 
The R\^ole of Rings}
\author{J.P. Wittmer$^{1}$\thanks{email:jwittmer@dpm.univ-lyon1.fr}, 
P.~van~der~Schoot$^{2}$, A. Milchev$^{3}$ and J.-L.~Barrat$^{1}$}
\address{
$^{1}$ D\'{e}partment de Physique des Mat\'{e}riaux, Universit\'{e} Lyon I\\
\& CNRS,\\
69622 Villeurbanne Cedex, France.\\
$^{2}$ Department of Applied Physics, Technische Univesiteit Eindhoven,\\
Postbus 513, 5600 MB Eindhoven,\\
The Netherlands\\
$^{3}$ Institute for Physical Chemistry, Bulgarian Academy of Sciences,\\
$1113$ Sofia, Bulgaria.}
\maketitle
\begin{abstract}
We investigate by means of a number of different dynamical Monte Carlo
simulation methods the self-assembly of equilibrium polymers in dilute,
semidilute and concentrated solutions under good-solvent conditions. In our
simulations, both linear chains and closed loops compete for the monomers,
expanding on earlier work in which loop formation was disallowed. Our
findings show that the conformational properties of the linear chains, as
well as the shape of their size distribution function, are not altered by
the formation of rings. Rings only seem to deplete material from the solution
available to the linear chains. In agreement with scaling theory, the rings
obey an algebraic size distribution, whereas the linear chains conform to a
Schultz--Zimm type of distribution in dilute solution, and to an
exponentional distribution in semidilute and concentrated solution. A
diagram presenting different states of aggregation, including monomer-,
ring- and chain-dominated regimes, is given. The relevance of our work in
the context of experiment is discussed.
\end{abstract}

\sloppy         

\centerline{\today}

\vskip 1.0truecm \centerline{PACS numbers: 82.35.+t, 61.25H, 64.60C } 

\section{Introduction.}\label{sec:Introduction}

Solutions of highly elongated, cylindrical giant micelles are arguably among
the best studied of the so-called equilibrium polymers \cite{com:GM}.
Equilibrium polymers are formed in reversible polymerisation processes, and
are in chemical equilibrium with each other, i.e., monomeric material is
continually exchanged between the assemblies. An aspect not at all well
understood is why ring closure seems to be unimportant in solutions of
linear micelles, although this would remove unfavorable free ends (``end
caps'') from the solution. Closed loops have been observed in electron
microscopic images of giant micelles \cite{Clausen92}, but --- as has been
argued elsewhere \cite{CC90,Porte,SW99} --- in too low concentrations to
significantly influence the properties of micellar systems. In other types
of equilibrium polymer, such as liquid sulfur, the presence of rings is on
the other hand thought to be all-important \cite{Petschek}. It is believed
that rings are suppressed in those self-assembled polymeric systems that are
sufficiently rigid on the scale of the individual monomers\cite{CC90,Porte,SW99}.

A useful model describing the self-assembly of giant micelles is what we
call the {\em Restricted} Model of equilibrium polymers, where {\em by
definition} ring closure and branching of chains are disallowed, and only
linear chains form. In this model, the linear self-assembly is regulated by
a free energy penalty associated with the free ends, the so-called end cap
(free) energy $E$. The end cap energy is normally presumed to be a constant
independent of the chain size or aggregation number, $N$, and of the overall
monomer density $\phi $\cite{com:LP}. The basic scaling predictions for the
equilibrium polymerisation within the Restricted Model \cite
{CC90,PvS97,Wheeler,Schaefer} are based on classical polymer physics\cite
{Degennesbook}, and have been tested by two of us (JPW, AM) by means of
various Monte Carlo approaches \cite{Potts,WMC98,MWL00}. 

Despite their inherent polydispersity, self-assembled linear polymers
resemble in many ways conventional polymers, that is, polymers with a fixed
molecular weight. Indeed, the known statistical properties of conventional
polymers have been applied quite successfully to predict the probability
distribution function of the size of equilibrium polymers. In agreement with
theoretical predictions \cite{CC90}, the size distribution function has for
instance been shown to decay essentially exponentially with size in the
semidilute and concentrated regimes, because then correlations generated by
the excluded volume interaction are small \cite{WMC98}. A typical aggregate
size distribution in a solution with strongly overlapping chains, obtained
by computer simulation, is given in Fig.~\ref{fig:cLRN} (open squares).

In the present work we relax the no-loop constraint, and discuss systems of
(mainly) {\em flexible} equilibrium polymers under good-solvent conditions,
where linear chains have to compete with rings for the available monomers at
given total density $\phi $, and end cap energy $E$. Branching of chains
remains forbidden. We discuss in detail the number density of chains and
rings of in different regimes (dilute {\em versus} strong overlap, ring 
{\em versus} chain dominated), and construct the complete ``phase'' diagram of
this {\em Unrestricted} Model. Following up an an earlier, less extensive study 
\cite{MWL00}, we report in this paper results obtained with a lattice-based
Monte Carlo method\cite{WMC98,BFM}, and with a more recent off-lattice Monte
Carlo scheme \cite{MWL00,OLMCold}. We have succeeded in {\em mapping} the
results of both methods onto each other, by making use of the natural length
and energy scales that describe the configurations of the equilibrium
polymers. This has enabled us to extract from the data the unknown
prefactors that enter in the standard scaling theory, allowing us to
construct a full diagram of states. Qualitatively, we confirm older, less
elaborate descriptions \cite{Porte,SW99}. 
As can be seen in Fig.~\ref{fig:cLRN}, 
the number density of rings as a function of aggregation number
(indicated by the filled circles) is strongly singular, and dominated by a
lower cut-off, that is, by the smallest ring allowed in the simulation.
Effectively, this causes the crossover between dilute and semidilute regime,
and that between ring- and chain-dominated regimes to coincide, at least for
large end-cap energies $E$.

This paper is organized as follows. We start in Sec.~\ref{sec:Algo} with a
brief presentation of the computational methods applied. Some technicalities
concerning the parameters used and the configurations sampled are considered
in Sec.~\ref{sec:Para}. The main computational results of this paper are
presented in the Sections~\ref{sec:Size},~\ref{sec:Lin} and \ref{sec:Ring}.
In Sec.~\ref{sec:Size} we discuss conformational properties, recall some
notions and concepts of the physics of conventional polymers (of fixed
size), and determine the size $\xi $ of the excluded volume blob \cite
{Degennesbook}. The following three Sections~\ref{sec:Mass},~\ref{sec:Lin}
and \ref{sec:Ring} are dedicated to the mass distributions. In the short
Sec.~\ref{sec:Mass} we fix some notions, refine the problem and pose
specific tasks. Subsequently, we discuss the number density of linear chains
of size $N$, $\mbox{$c_1$}(N)$, in Sec.~\ref{sec:Lin}. Results obtained
within the Restricted Model (with rings suppressed) and the Unrestricted
Model (with rings allowed) are compared. We attempt to map both Monte Carlo
methods onto each other, and discuss the physics at extremely high
densities, where our off-lattice Monte Carlo approach exhibits features
linked to packing effects, not present in the lattice-based Monte Carlo
method. The number density of rings of aggregation number $N$, 
$\mbox{$c_0$}
(N)$, and its connection to the size and mass distribution of the linear
chains are discussed in Sec.~\ref{sec:Ring} -- the core of the paper. In
Sec.~\ref{sec:PD} we calculate the different regimes for systems of flexible
equilibrium polymers in a good solvent, and compare this with the
``measured'' distribution of monomers over the rings and linear chains. In
the final Sec.~\ref{sec:Discussion} we summarize our findings, and speculate
on the relevance of ring formation in some recent experiments 
\cite{Schurtenberger,Folmer,Greer}.

\section{Algorithms.}\label{sec:Algo}

All the computational algorithms we discuss are standard Monte Carlo schemes
which have already been described and tested in depth in recent publications 
\cite{Potts,WMC98,MWL00}. We shall therefore be extremely brief.

The first method which was introduced to simulate the equilibrium polymers
within the Restricted Model, is a grand canonical lattice algorithm based on
a mapping on a Potts model\cite{Potts}. Because this model lives on a simple
cubic lattice, closed loops are by geometry forced to consist of an even
number of monomers. Results obtained with this model will only be briefly
mentioned (in connection with Fig.~\ref{fig:cLRN}). We discuss in more
detail results obtained with two canonical Monte Carlo approaches --- a
lattice scheme \cite{WMC98} based on the Bond Fluctuation Model (BFM) \cite
{BFM}, and a more recent off-lattice Monte Carlo (OLMC) approach \cite{MWL00}
generalizing an efficient bead-spring model. All BFM and OLMC simulations
have been done in three spatial dimensions, 
the Potts model simulations in two and three dimensions.

The BFM used in the present investigation is athermal,
besides the {\em constant} scission energy $U_{bond}=-J$, which
characterizes the bonded interaction: an energy $J>0$ is released every time
a bond forms. Apart from the no-overlap conditions modelling hard-core
excluded volume interactions, no other bonded (such as a stiffness
potential) or non-bonded interaction has been used in the present study,
though they might be readily included.

In the off-lattice model, beads interact via a so-called (shifted) FENE
potential for the bonded, and a purely repulsive Morse potential for the
non-bonded interactions \cite{OLMCold}. The bonded interaction 
\begin{equation}
U_{bond}(r)=-K(r_{max}-r_{o})^{2}\ln \left[ 1-\left( \frac{r-r_{0}}
{r_{max}-r_{0}}\right) ^{2}\right] -J
\label{eq:FENE}
\end{equation}
depends non-trivially on the distance $r$ between two monomers. Here, $J$
denotes again the constant part of the scission energy, and $K$ the spring
constant. The former quantity will be related with the model-independent effective
end-cap free energy $E$, in Sec.\ref{sec:Lin}. Note further that the FENE
potential is harmonic near its minimum at $r_{0}$, while exactly at $r=r_{0}$, 
$U_{bond}=-J$. The potential diverges logarithmically to infinity if 
$r\rightarrow r_{max}$ and if $r\rightarrow r_{min}=2r_{0}-r_{max}$, where 
$
r_{\max }$ and $r_{\min }$ are the maximum and minimum extension of the
spring. Following ref.\cite{OLMCold}, we set $r_{max}=1$, 
$r_{max}-r_{0}=r_{0}-r_{min}=0.3$ and $K=40$. Units are chosen such that 
$k_{B}T=1$.

Both models use local jump attempts. The time is measured, as usual, in
Monte Carlo steps (MCS) {\em per monomer}. Every monomer is chosen at
random, and allowed to perform a move subject to a Metropolis acceptance
probability \cite{WMC98,MWL00}. In simulations the bonds between neighbors
along the backbone of a chain are constantly subject to scission and
recombination events. Since chains are only transient objects the data
structure of the chains can only be based on the individual monomers, or,
and this is the way we operate, on the notion of saturated and unsaturated
bonds. An unsaturated bond does not connect a monomer to another one, a
saturated bond does. Hence, we use no direct chain information, but lists of
pointers linking up the bonds\cite{WMC98}. 
With a given frequency one of the bonds (saturated or
unsaturated) is chosen at random. If that bond happens to be saturated an
attempt is made to break it, if it is unsaturated, i.e., if the
monomer is at the end of a chain or a free monomer, an attempt is made to
create a bond with another monomer sufficiently close. Obviously, for
reasons of detailed balance, the bond formed must come out of the same set
or range of bonds out of which also scission events are allowed to take
place \cite{WMC98,MWL00}. We do not allow for branching in the present
study. The minimal length of a closed loop is $N_{c}=3$, i.e. closed dimers
are not permitted.
Free monomers are also not allowed to self-saturate their two unsaturated bonds. 

\section{Parameters and configurations.}\label{sec:Para}

The only two model parameters of operational relevance in the present study
to tune the system properties are the scission energy $J$ and the number
density $\phi $. The starting configurations consist in both sets of
simulations of randomly distributed and non-bonded monomers, which we cool
down step by step (a sequence of so-called `T-Jumps'), each step sampling a
higher scission energy up to a maximum $J=15$. This was done in order to
produce a sufficient chain length and concentration variation to be able to
put to the test the theoretical scaling predictions.

Due to the constant breaking and recombining of the bonds, equilibration is
much faster in our equilibrium polymeric system than is usually observed in
systems of conventional polymers with fixed bonds. The algorithm presented
above (using the pointer lists between bonds) allows us to simulate a large
number of particles at very modest expenses of operational memory.

In our BFM simulations, we varied the number density over
three orders of magnitude from $\phi =0.000125$ ($1,000$ monomers per box)
to $\phi =0.075$ (containing $75,000$ monomers per box). For densities
smaller than $\phi =0.0125$ cubic lattices with volume $200^{3}$ were used,
whilst for higher densities a smaller box sufficed of volume $100^{3}$. Note
that in the BFM every monomer consists of an elementary
cube, whose eight sites on the cubic lattice are blocked for further
occupation\cite{BFM}. As a result of this, the volume fraction of material
is given by $8\phi $. As was shown elsewhere\cite{Wolferl,MWC00}, volume
fractions of around $8\phi \approx 0.5$ are already quite dense within the
BFM. The reason is that at higher densities than this,
the system turns glassy due to the blockage of neighboring sites by other
monomers. 

Most of the off-lattice Monte Carlo results involve $65,536$ particles for
number densities between $\phi =0.0325$ and $\phi =2$ (and appropriately
chosen box sizes). Generally, we have sampled with the off-lattice Monte
Carlo method more systems in the high and extremely high density regime,
while we have explored with the BFM more systems in the
dilute and semidilute regimes. Note that the highest densities in our
off-lattice Monte Carlo simulations correspond to extraordinarily
concentrated solutions. Indeed, if we estimate the corresponding volume
fractions, we find that in our simulations these vary between $0.0072$ and 
$0.44$. (To obtain this estimate, an effective bead volume $v\approx \pi
l^{3}/6\approx 0.22$ was used, with a measured mean bond length of $l\approx
0.75$.) The latter value of $0.44$ has to be compared with the (only
slightly larger) hard-sphere freezing volume fraction (``Alder transition'')
of about one half. Our simulations therefore extend to the ``melt'' regime
of a dense liquid. 

In passing we note that strictly speaking the mean bond length $l$ is not a
constant, but decreases weakly with increasing density. However, this does
not pose a serious problem, for we find that the total interaction energy
per bond does remain roughly constant, with a $U_{total}\approx -(J-0.4)$
for all overall monomer densities $\phi $ and scission energies $J$ probed. 

Periodically, the whole system is examined and various moments and
distributions, such as the number densities $c_{i}(N)$, are counted and
stored. Here, and below, we refer to ring-related quantities by using a
subscript $i=0$, and to chain-related ones by a subscript $i=1$. The number
densities $c_{i}(N)$ are normalized such that $\phi _{i}=\sum_{N}Nc_{i}(N)$,
with $\phi _{i}$ the overall density of monomers in species $i$. Obviously,
the sum of monomers in rings, \mbox{$\phi_0$},\ and that in linear chains, 
\mbox{$\phi_1$},\ is equal to the overall monomer density $\phi =\phi _{0}+\phi
_{1}$. (Free monomers are counted as linear chains of length $N=1$.)
Obviously, $\phi =\mbox{$\phi_1$}$ and $\mbox{$\phi_0$}=0$ in the Restricted
Model, where rings are disallowed. We emphasize that the mean chain length 
$\left\langle N\right\rangle \equiv \sum_{i,N}Nc_{i}(N)/\sum_{i,N}c_{i}(N)$ \
remains always two orders of magnitudes smaller than the total particle
number within the box. From our previous studies within the Restricted Model 
\cite{WMC98,MWL00}, we expect finite box-size effects to be small.

Because of the differences in the lattice and off-lattice algorithms, we
cannot expect the results obtained using these two methods to be directly
comparable, even when the same parameters $(J,\phi )$ are used. To make a
comparison possible, all relevant system parameters have to be mapped onto
each other. This we do by taking the dilute limit as reference state. As the
intrinsic energy scale we use $E=J+\delta J$, with a model-dependent shift
parameter $\delta J$ presented in Sec.\ref{sec:Lin}. The intrinsic length
scales are the mean bond length $l$, and a length $l_{p}$ to be discussed in
the next section.

\section{Distributions of size.}\label{sec:Size}

Let us first discuss the conformational properties of equilibrium polymers,
and show that not only they follow the same universal laws as conventional
polymers, but also that there is no essential difference in the behaviour of
linear chains in the Restricted and Unrestricted Models.

\begin{figure}[tbp]
\centerline{\epsfig{file=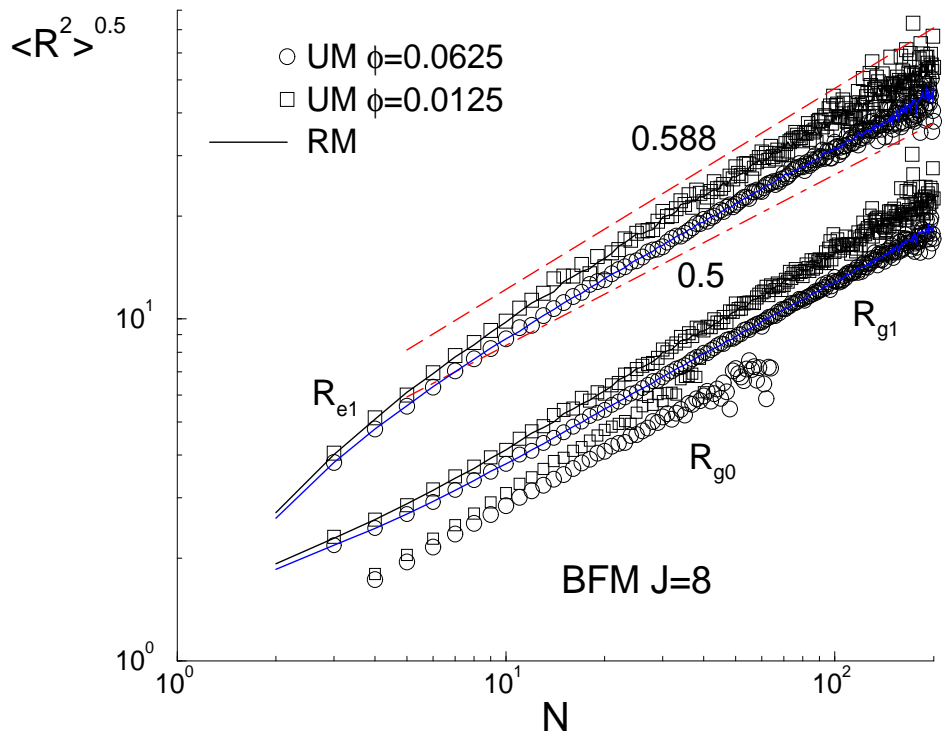,width=120mm,height=100mm}}
\caption{
Mean chain size $\langle R(N)^2 \rangle^{0.5}$ for given aggregation number
$N$, as obtained from
the Restricted Model (RM -- lines) and Unrestricted Model (UM -- symbols)
BFM simulations, at two different
densities at $J=8$. }
\label{fig:RJ8N}
\end{figure}

The configurational behaviour of the equilibrium polymers is most easily
demonstrated by plotting their mean size as a function of the aggregation
number $N$, as is done in Fig.\ref{fig:RJ8N}. Indicated are the mean
end-to-end distance of the linear chains, $R_{e1}(N)$, and the radii of
gyration of the chains and the rings, $R_{g1}(N)$ and $R_{g0}(N)$. Averages
have been taken over all linear chains ($i=1$) or rings ($i=0$) of 
{\em given} mass $N$ in the simulation box. Only results obtained with the bond
fluctuation method are shown, for two different densities and a fixed value
of the scission energy $J=8$. Similar results have been obtained with the
off-lattice Monte Carlo method. Symbols are used for the distributions from
the self-assembled chains in the presence of rings, lines for those
without rings. Despite the fact that the simulations point at the presence
of a large number of rings within the Unrestricted Model, with the dilute
systems even being ring dominated, there is no measurable difference between
the results for the linear chain sizes in the Restricted and Unrestricted
Model calculations. This we quite generally find for all densities probed,
and within both simulation methods. 

Indicated in the figure are two dashed lines, giving the theoretical slopes
valid for very long polymers in dilute and concentrated solution \cite
{Degennesbook}. In the dilute limit the chains are swollen and their size $R$
is described by the scaling relation $R\propto N^{\nu}$, with a
self-avoiding walk exponent of $\nu =0.588$. Surprisingly, the relatively
small rings shown in Fig.\ref{fig:RJ8N} also closely follow the (asymptotic)
scaling behaviour of their linear counterparts. In the following, the dilute
limit will be used as reference state to be able to make a comparison
between the different simulation models. As natural lengths we use the mean
bond length $l$ and an effective bond length $b=b_{e1}$, the latter
determined from a fit of the simulation data to the scaling relation 
$R_{e1}(N)=b_{e1}N^{0.588}$ . These two lengths in turn suggest a measure for
the stiffness of the chains, which we shall call the ``persistence length'' 
$l_{p}=b/l$ of the chains, although obviously it is a dimensionless quantity.
For the BFM we find $b\approx 3$ \cite{WMC98}, and for
the off-lattice Monte Carlo method $b\approx 0.92$. In both cases this gives
a persistence length $l_{p}=b/l$ that are of similar value ($l=2.733$, 
$l_{p}\approx 1.2$ for the BFM; $l=0.758$, $l_{p}\approx
1.1$ for off-lattice model). The found values for $b$ are essentially
identical to what was obtained previously for monodisperse linear polymers 
\cite{OLMCold,MWC00}. In a similar fashion, one may define and measure the
prefactors $b_{g0}$ and $b_{g1}$ from the radii of gyration $R_{g0}$ and 
$R_{g1}$. In line with our previous work on conventional polymers studied in 
\cite{MWC00}, we find $b_{g0}/b\approx 0.3$ and $b_{g1}/b\approx 1/\sqrt{6}$. 
This result holds again for both simulation methods.

Figure \ref{fig:RJ8N} shows that, as expected\cite{Degennesbook}, the
excluded volume correlations in our equilibrium systems are screened out for
strongly overlapping chains. More precisely, the chains become Gaussian
chains of blobs of size $\xi (\phi )=b_{g1}g^{\nu }$, each blob containing 
$g$ monomers, with 
\begin{equation}
g=g(\phi )=G\left( b^{3}\phi \right) ^{-1/(3\nu -1)}
\label{eq:gblob}
\end{equation}
and G an as yet unknown prefactor. The value of $G$ may be fixed using the
classical definition for the crossover density of a monodisperse solution: 
$4\pi/3 \ \xi^3\phi = g$.
This definition gives for the prefactor $G\approx 5.2$, a value that turns
out to be numerically consistent with alternative estimates obtained from
fits to the scaling relations $R_{e1}(N)=b_{e1}g^{\nu }(N/g)^{1/2}$ or 
$R_{g1}(N)=b_{g1}g^{\nu }(N/g)^{1/2}$.

\begin{figure}[tbp]
\centerline{\epsfig{file=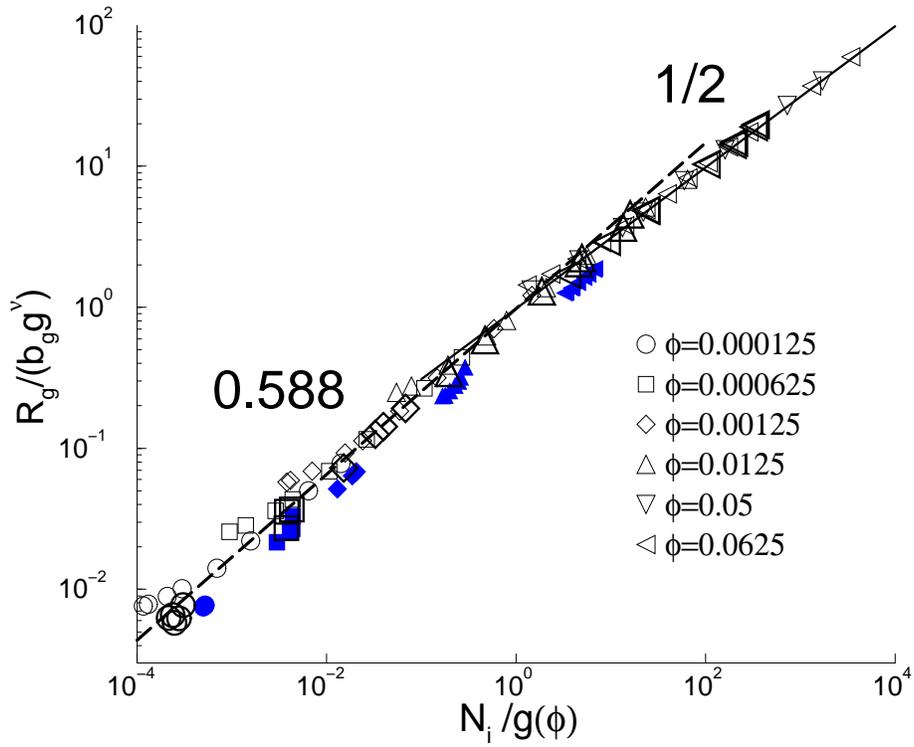,width=120mm,height=100mm}}
\caption{
Scaling plot of the scaled radius of gyration
$\mbox{$R_g$}/\xi(\phi)$ versus scaled aggregation number $N_i/g(\phi)$, for
rings (filled symbols) and
linear chains (small symbols for RM, large open symbols for UM).
$N_i$ is here the mean aggregation number of the respective quantity
presented. All data shown are BFM results.
\label{fig:Rgscal}}
\end{figure}

A further test of the configurational properties of the equilibrium polymers
is provided by determining the dependence of the first moment of the size
distributions on the mean aggregation number $\left\langle
N_{i}\right\rangle \equiv \sum_{N}Nc_{i}(N)/\sum_{N}c_{i}(N)$. Our results
are presented in Fig.\ref{fig:Rgscal}, where we have plotted the radius of
gyration divided by the blob size $y=R_{gi}/\xi =R_{gi}/(b_{g1}g^{\nu })$, 
{\em versus} the mean aggregation number divided by the number of monomers
per blob $x=\left\langle N_{i}\right\rangle /g(\phi )$. In order to
calculate these reduced quantities, we used eq.(\ref{eq:gblob}) together
with our previous estimate for $G=5.2$. The figure shows that the mean chain
sizes of the living polymers again follow the same universal master curve as
conventional polymers \cite{Degennesbook}. Again we observe that the ring
sizes, indicated by the full symbols, follow the same scaling law as the
linear chains, which is actually rather surprising considering their
relatively small size. The fact that the two slopes that indicate the scaling
behaviour expected in the dilute and semidilute regimes cross nicely at 
$(x,y)\approx (1,1)$, shows that our estimate of $G$ is actually rather
accurate. Alternatively, we could have used this crossover point between the
two regimes to define the prefactor $G$.

In conclusion, the conformational properties of equilibrium chains are
(within numerical accuracy) identical to those of polymers of fixed length.
This is in contrast to earlier speculation in the literature, where it was
surmised that rings might strongly influence the screening of excluded
volume close to the crossover from the dilute to the semidilute regime\cite{CC90}. 
We find that the universal functions are neither altered by the
polydispersity, nor by the presence of rings. The blob size $\xi $ is a
function of total monomer density $\phi $ only.

\section{Mass distributions of equilibrium polymers.}\label{sec:Mass}

Next we turn our attention to the probabilities of finding aggregates of a
certain aggregation number. For equilibrium polymers that are long compared
to the effective bond length $b$, the main departure from conventional
theory of flexible polymer solutions is that the reversibility of the
self-assembly process ensures that the degrees of polymerisation are in
thermal equilibrium. This means that the distributions $\mbox{$c_0$}(N)$ and 
$\mbox{$c_1$}(N)$ are not fixed, but minimize the thermodynamic potential 
$\Omega [\mbox{$c_0$}(N),\mbox{$c_1$}(N)]$ of the system. It is natural to
attempt a simplified theoretical description, using a Flory-Huggins type of
mean-field approximation 
\begin{equation}
\Omega [\mbox{$c_0$}(N),\mbox{$c_1$}(N)]=\sum_{i=0}^{1}\sum_{N=1}^{\infty
}c_{i}(N)\left( \log (c_{i}(N)l^{3})+\mu N+\tilde{f}_{i}\right) 
\label{eq:MFhyp}
\end{equation}
where we have written the thermodynamic potential as a sum over the
different species $i=0$ for rings and $i=1$ for linear chains, and over all
possible aggregation numbers $N$. The factor $l^{3}$ in the logarithm enters
for dimensional reasons, where we set $l$ equal to the mean bond length of
the chains in the dilute regime; all energy units are measured in units of
thermal energy, $k_{B}T=1$. This ansatz for the thermodynamic potential is
strongly motivated by its success describing the properties of equilibrium
polymers within the Restricted Model, where ring formation is suppressed 
\cite{WMC98,MWL00}. The first term on the right is the usual translational
entropy. The second term represents a Lagrange multiplier or chemical
potential which fixes the total monomer density 
$\phi =\mbox{$\phi_0$}(\mu )+\mbox{$\phi_1$}(\mu )$. 
All contributions to the free energy which are extensive or linear in $N$
are absorbed in this Lagrange multiplier. The as yet not specified terms 
$\tilde{f}_{i}$ describe the free energy contributions not extensive in the
degree of polymerisation of the rings and linear chains. In general these
may depend on the interactions between different chains and chain parts, and
as a rule differ in the dilute, semidilute and melt regimes. We stress that
in the prescription of eq.~(\ref{eq:MFhyp}) terms (such as virial terms)
which are not conjugate to $\mbox{$c_0$}(N)$ or $\mbox{$c_1$}(N)$ need not
be made explicit, as they are absorbed in $\mu $. Inspired by the results of
the previous section, the central assumption of eq.~(\ref{eq:MFhyp}) is that
the two distribution functions are only coupled via the chemical potential
that makes sure that the total amount of monomers is conserved.

In equilibrium, the distribution functions functionally minimise the
thermodynamic potential, $\delta \Omega /\delta c_{i}=0$, giving 
\begin{eqnarray}
l^{3}\mbox{$c_0$}(N) &=&\exp (-\mbox{$f_0$}(N,b^{3}\phi ,l_{p})-\mu N)
H(N-\mbox{$N_c$}(l_{p})) \\  \label{eq:cLRN}
l^{3}\mbox{$c_1$}(N) &=&\exp (-E(l_{p})-\mbox{$f_1$}(N,b^{3}\phi )-\mu N)
\nonumber  
\end{eqnarray}
where we set $\tilde{f}_{0}+1=\mbox{$f_0$}(N,b^{3}\phi ,l_{p})$ and 
$\tilde{f}_{1}+1=E(l_{p})+\mbox{$f_1$}(N,b^{3}\phi )$ for convenience. 
The free energy associated with the linear chains $\tilde{f}_{1}$ is split into a
model- and density-invariant end-cap free energy $E=J+\delta J$ (containing
the operational scission energy $J$ and a shift factor $\delta J$ discussed
below), and a remaining part that somehow describes excluded volume
correlations \cite{Degennesbook}. The Heaviside function $H(x)$ enforces a
smallest possible ring $\mbox{$N_c$}(l_{p})$, in effect a lower cut-off. In
actual systems this cut-off may depend on factors such as the detailed
chemistry of the equilibrium polymers in hand, and on their bending
stiffness not explicitly modelled here.

Before we are able to comprehensively analyse our computer simulation
results aided by eq.~(\ref{eq:cLRN}), our task is (1) to complete the mapping
of the simulation methods already started in the previous section, (2) to
show that the Flory-Huggins ansatz is indeed justified, and that the ring
and chain distributions decouple, (3) to identify the free energy terms 
\mbox{$f_0$} \ and \mbox{$f_1$}, (4) to show that the $f_{i}$ are only
functions of $N$ and the total density $b^{3}\phi $ as indicated, and (5) to
determine the equilibrium values of $\mbox{$\phi_0$}/\phi $, and 
$\left\langle N_{i}\right\rangle $ for a given system $(J,\phi )$. In the
next two sections~\ref{sec:Lin} and \ref{sec:Ring} we consider the
distribution functions of the linear chains and rings, and analyse the free
energy terms $f_{i}$ (tasks 3 and 4). The mapping (task 1) will be completed
in Sec.~\ref{subsec:LinDilute}, where we relate the natural (and in
principle measurable) energy scale $E$ with the operational parameter $J$.
We show there that the distribution functions of linear chains in the
Unrestricted Model can be computed from those in the Restricted Model, if
the density\ of monomers in chains \mbox{$\phi_1$} is given (task 2). In
Sec.~\ref{subsec:Linscal} we study the density crossover scaling for 
$\mbox{$N_1$}(\mbox{$\phi_1$},\phi ,E)$. The relative distribution of
monomers in rings and in linear chains (task 5) will be considered in 
Sec.~\ref{sec:PD}.

\section{Mass distributions of linear chains.}\label{sec:Lin}

In this section we focus attention on the size distributions of the linear
equilibrium polymers, as determined by computer simulation, within both the
Restricted and Unrestricted Model settings. 
A typical example is given in Fig.\ref{fig:cLRN}, where results are shown
obtained with the off-lattice algorithm at high densities, that is, in the
limit of strongly overlapping chains ($\phi =1.5$).
The distribution functions for the Restricted and Unrestricted Models are
both to high accuracy pure exponentials. Under conditions where linear
chains dominate in the Unrestricted Model ($\mbox{$\phi_1$}\approx \phi $),
we find that results from the Restricted and Unrestricted Models converge
systematically. This is generally true in the limit of strongly overlapping
equilibrium polymers, because then linear chains dominate the population of
equilibrium polymers. (See also below.) At variance with recently published
results of a simulation study \cite{Yannick}, we do not observe any sign of
an algebraic singularity in the distribution function of the linear chains,
even at the extremely high volume fraction of $\phi =2$. Results (not shown)
from the BFM simulations confirm all trends observed
with the off-lattice Monte Carlo treatment.

\begin{figure}[tbp]
\centerline{\epsfig{file=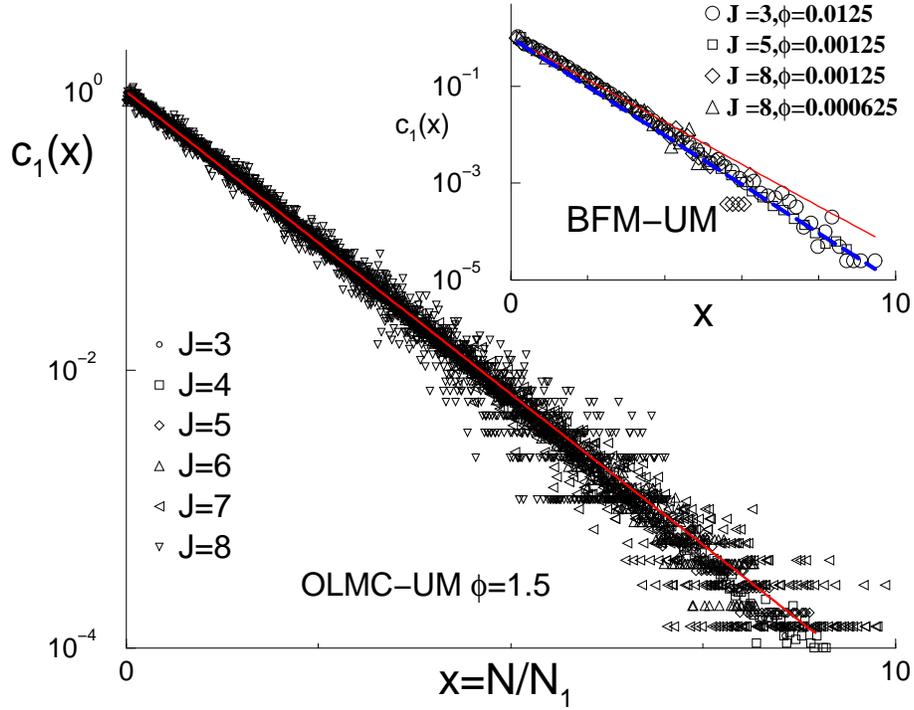,width=120mm,height=100mm}}
\caption{
Normalized distributions for linear chains $\mbox{$c_1$}(x)$ versus the
scaled aggregation number
$x=N/\mbox{$N_1$}$, where again \mbox{$N_1$} denotes the mean aggregation
number of the linear chains. Main figure: Data collapse on the exponential
distribution $\mbox{$c_1$}(x)=\exp(-x)$
for the OLMC data within the UM, at $\phi=1.5$. 
Inset: Similar plot showing the BFM data in the dilute limit. 
As for the RM\protect\cite{WMC98,MWL00},
we find $\mbox{$c_1$}(x)\propto \exp(-\mbox{$\gamma$} x)$ with
$\mbox{$\gamma$}
\approx 1.16$ a critial exponent. }
\label{fig:pLR1}
\end{figure}

A more critical test of our statement that rings only marginally influence
the distribution functions of the linear chains is provided in Fig.\ref
{fig:pLR1}, we have plotted the distribution of linear chains 
$\mbox{$c_1$}(N)$ against the natural scaling variable $x=N/\left\langle
N_{1}\right\rangle $. Apparently, in the regime of strongly overlapping
chains (main figure) the data collapse onto $\mbox{$c_1$}(N)\propto \exp
\left( -N/\left\langle N_{1}\right\rangle \right) $ for a whole host of
scission energies $J$. (Note that chain statistics deteriorate in the tail
of the distribution.) In the dilute limit we find a similar data collapse,
but with slightly different slope, pointing at a distribution of the type 
$\mbox{$c_1$}(N)\propto \exp \left( -\gamma N/\left\langle N_{1}\right\rangle
\right) $, with $\gamma $ a constant identified below as a critical
exponent. See the inset of figure \ref{fig:pLR1}, where results obtained
from the BFM simulations covering the dilute limit are
shown. Identical distributions have been found in both limits for systems
without rings \cite{WMC98,MWL00}. 

How can our observations be understood? We have demonstrated in a previous
publication that within the Restricted Model, the free energy term 
\mbox{$f_1$} in the expression for the distribution function of linear
chains in eq.~(\ref{eq:cLRN}) can be calculated from the scaling theory of
conventional polymers in a good solvent \cite{Degennesbook}. Agreement with
simulation data turned out to be remarkably good \cite{WMC98,MWL00}. 
We show here
that this ansatz is also justified when rings are present, as long as the
overall monomer density is not extremely high (Sec.~\ref{subsec:LinMelt}).

\subsection{The Mean-Field solvent.}\label{subsec:LinMF}

Starting with the (hypothetical) conditions where mean-field type of
behaviour dictates the distribution functions, \mbox{$f_1$} \ is a constant. 
\mbox{$f_1$} may then be absorbed in $E$, i.e., in that case we may simply
put $\mbox{$f_1$}=0$. From eq.~(\ref{eq:cLRN}) we thus read off that within
mean-field theory, the distribution function must be a pure exponential 
$l^{3}\mbox{$c_1$}(N)=\exp (-E-\mu N)$. It is easily shown that for this type
of distribution, the equality $\mu =1/\left\langle N_{1}\right\rangle $
holds, albeit exactly only in the limit of large mean aggregation numbers 
$\left\langle N_{1}\right\rangle \gg 1$. For the mean-chain length we obtain 
\begin{equation}
\left\langle N_{1}\right\rangle =Al^{3\eta }\mbox{$\phi_1$}^{\eta }\exp
(\delta E)
\label{eq:NLgeneric}
\end{equation}
with an amplitude $A=\exp (1/2),$ and mean-field exponents $\eta =\delta =1/2
$. Note that $\left\langle N_{1}\right\rangle $ in the Unrestricted Model is
a function of $\phi _{1}=\phi _{1}(\phi ,E)$,\ and only indirectly dependent
on the control parameter $\phi $.

\subsection{The dilute limit.}\label{subsec:LinDilute}

Although largely unimportant in the semi-dilute and concentrated regimes,
correlation effects do matter in the dilute regime, as we have in fact
already seen in our discussion of the conformational properties in Sec.~\ref
{sec:Size}. Using the known statistical properties of self-avoiding walks,
we infer that the free energy of a chain $\mbox{$f_1$}$ decreases
logarithmically with the degree of polymerisation $N$ \cite{Degennesbook} 
\begin{equation}
\mbox{$f_1$}=-(\gamma -1)\log (N).
\label{eq:fLdilute}
\end{equation}
where $\gamma \approx 1.158$ is the susceptibility exponent of the 
$n\rightarrow 0$ vector model in 3 spatial dimensions\cite{CCP98}. 
Eq.~(\ref{eq:fLdilute}) leads to a Schultz-Zimm distribution for the linear chains 
\begin{equation}
l^{3}\mbox{$c_1$}(N)=N^{\gamma -1}\exp (-E-\mu N).
\label{eq:Schultz-Zimm}
\end{equation}
which indeed is born out by our simulation results, presented in the inset
of Fig.~\ref{fig:pLR1} \cite{com:depletion}, since in the dilute limit we
have $\mu =\gamma /\left\langle N_{1}\right\rangle $ provided $\left\langle
N_{1}\right\rangle \gg 1$ \cite{CC90,WMC98}. 

As already advertised in the previous section, we use the simulation results
obtained in the dilute limit to map the different simulation models onto
each other. The natural energy scale $E=J+\delta J$ is established by
fitting the data for $\mbox{$c_1$}(N)$ to eq.~(\ref{eq:Schultz-Zimm}), using
the equality $\mu =\gamma /\left\langle N_{1}\right\rangle $. Having done
this for the Restricted and Unrestricted Model simulation at various values
of $J$ and $\phi $, we find $\delta J=1.6$ for the BFM 
and $\delta J=1.7$ for off-lattice model. We stress that $\delta J$ is model
dependent, absorbing the different physical behaviour of the models on a
microscopic scale (see Sec.\ref{sec:Algo}). It can only be coincidental that
for the models we used the $\delta J$'s turn out to be almost identical.
Whatever the reason, the fixing of $\delta J$ completes the mapping.

The concentration dependence of the mean aggregation number in the dilute
regime, for future reference denoted by $N_{d}$, is given by 
\begin{equation}
N_{d}\equiv \left\langle N_{1}\right\rangle =\gamma /\mu =A_{d}l^{3\eta }
\mbox{$\phi_1$}^{\eta }\exp (\delta E)
\label{eq:NLdilute}
\end{equation}
with the exponents $\eta =\delta =1/(\gamma +1)$, and a prefactor 
$A_{d}=\gamma \Gamma ^{\delta }(\gamma +1)\approx 1.2$, in terms of the
exponent $\gamma $ and the usual gamma function $\Gamma $. Our simulations
confirm these exponents and the prefactor (results not shown).

\subsection{Semidilute solutions.}\label{subsec:LinSemi}

The aggregation number dependent free energy contribution 
eq.~(\ref{eq:fLdilute}) describes only dilute chains, {\em i.e.}, chains which are
too short to overlap. From the standard theory of conventional polymers \cite
{Degennesbook}, one expects excluded volume effects to be screened out when
the chains strongly overlap. Even then, this happens only for chains larger
than the blob size, that is, for aggregation numbers $N\gg g(\phi )$, where 
\mbox{$f_1$} \ levels off to
\begin{equation}
\mbox{$f_1$}=-(\gamma -1)\log (g_{1}(\phi ))=f_{s}+\frac{\gamma -1}{3\nu -1}
\log (b^{3}\phi ).
\label{eq:fLsemi}
\end{equation}
and $g_{1}(\phi )\propto $ $g(\phi )$; the quantity $g_{1}$ scales like 
$g$ (see eq.~\ref{eq:gblob}), but with the different prefactor 
\begin{equation}
G_{1}=\exp (-f_{s}/(\gamma -1))\approx 0.43
\label{eq:G1}
\end{equation}
fixed by a constant $f_{s}=0.13$, estimated below.

\begin{figure}[tbp]
\centerline{\epsfig{file=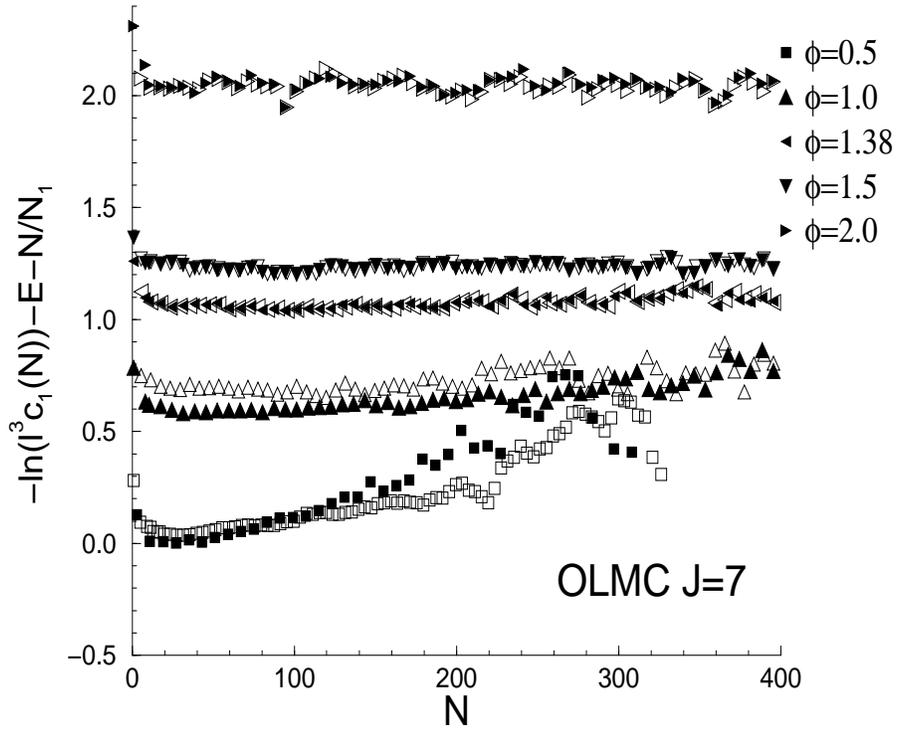,width=120mm,height=100mm}}
\caption{
Density crossover of the quantity $-\log(\mbox{$c_1$}(N)l^3)-E-N/
\mbox{$N_1$}$, obtained for $J=7$ from the Restricted (open symbols) and
Unrestricted Model (filled symbols) OLMC simulations. For conditions of
strong chain overlap, this quantity becomes chain length
independent and equal to the linear chain free energy $\mbox{$f_1$}(\phi)$
discussed in the main text. }
\label{fig:fLN}
\end{figure}

That \mbox{$f_1$} indeed switches from length dependent (dilute) to density
dependent (semidilute and melt) is shown in Fig.~\ref{fig:fLN}, where we
plot measured values of $\mbox{$f_1$}(N)$, using the mean-field relation 
$l^{3}\mbox{$c_1$}(N)=\exp (-E-\mu N)$ and $\mu =1/\left\langle
N_{1}\right\rangle $. Figure \ref{fig:fLN} shows that \mbox{$f_1$} \ is a
constant of the degree of polymerisation at high overlap concentrations 
\cite{com:fNlarge}, as it should. It explains the scaling relation 
observed in Fig.~\ref{fig:pLR1}, not least because the values of 
$\mbox{$f_1$}$ found for the Restricted and Unrestricted Models turn out to
be identical if the overlap is strong enough. At low densities ($\phi \leq
0.5$), \mbox{$f_1$}\ becomes dependent on the chain length, due to the
unscreened excluded volume interactions, but also on whether in the model
ring formation is allowed or not. In the Unrestricted Model ring formation
is allowed, and as a result of that some of the available material is stored
in rings. This causes a shift in the distribution of linear chains relative
to that in the Restricted Model. 
The trends of figure \ref{fig:pLR1} confirm this.

\begin{figure}[tbp]
\centerline{\epsfig{file=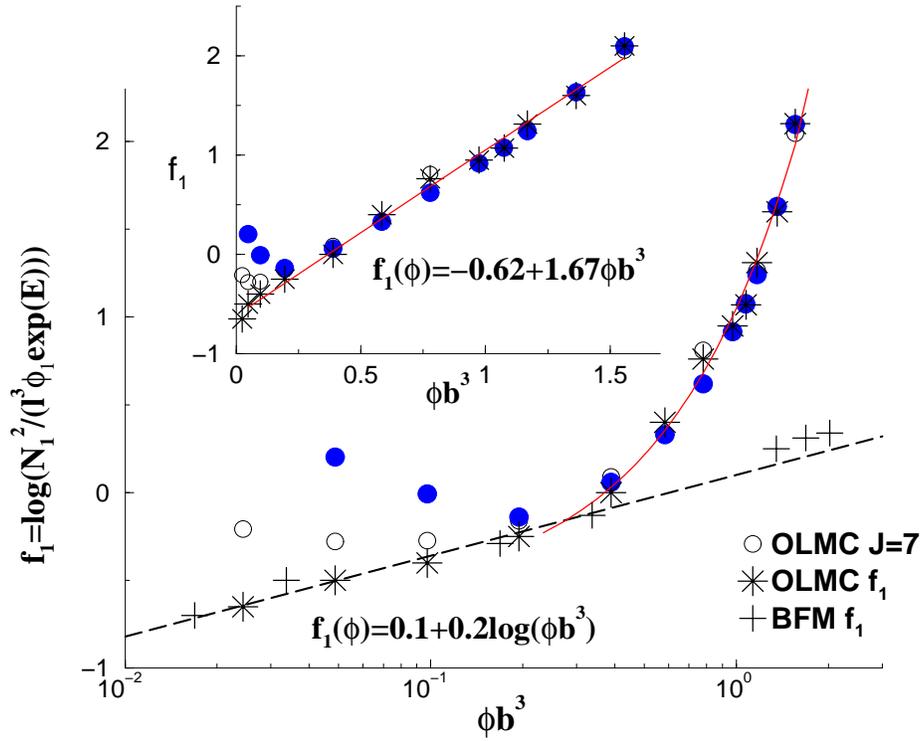,width=120mm,height=100mm}}
\caption{
Scaled aggregation number of the linear chains \mbox{$N_1$}\ versus the
dimensionless concentration $\phi b^3$ from OLMC simulations with the
scission energy put at $J=7$.
Values from both the Restricted (open circles) and Unrestricted Model OLMC
simulations (filled circles) are included. Also shown are the (asymptotic)
values, obtained independently from $\mbox{$c_1$}(N)$ for the largest $J$
available for a given density for OLMC (asterisks) and BFM (pluses).
At higher melt densities $\phi \geq 0.5$, we find
for the OLMC data a nonalgebraic dependence of $\phi$. This can be
clearly seen in the inset where we fitted the data to a function
$\mbox{$f_1$}(\phi)= B_0 + B_1 \phi b^3$,
with $B_0\approx -0.62$ and $B_1 \approx 1.67$. }
\label{fig:N1phi}
\end{figure}

Using the measured plateau values of $\mbox{$f_1$}$, we verify that the
density variation predicted by eq.~(\ref{eq:fLsemi}) holds, and estimate the
prefactor $G_{1}$ {\em via} eq.~(\ref{eq:G1}). 
This is done in Fig.~\ref{fig:N1phi}, 
where we have plotted the plateau values \mbox{$f_1$} \ versus
the dimensionless overall concentration of monomer, $b^{3}\phi $. Data from
the BFM (pluses) and the OLMC (asterisks) are included in the figure. 
Also shown are $\mbox{$f_1$}$ values as they may
be measured {\em via} the mean aggregation number in the strong overlap
limit (SOL), that is, in semidilute and concentrated solution 
\begin{equation}
N_{SOL}\equiv \left\langle N_{1}\right\rangle =l^{3/2}\mbox{$\phi_1$}^{1/2}
\exp \left( \frac{1}{2}E+\frac{1}{2}\mbox{$f_1$}(b^{3}\phi )\right).
\label{eq:NLSOL}
\end{equation}
which \ depends explicitly on both $\phi $ and $\mbox{$\phi_1$}(\phi ,E)$.
We use the directly measured $\langle N_1 \rangle$ \ and 
\mbox{$\phi_1$} \ to obtain $\mbox{$f_1$}(b^{3}\phi)$. 
This procedure gives identical results as the 
\mbox{$f_1$} \ obtained straight from the distribution functions. Again,
there is no observable difference between the results obtained with and
without ring formation. In line with the prediction of eq.~(\ref{eq:fLsemi}
), we find a logarithmic density dependence (dashed line) with a prefactor 
$(\gamma -1)/(3\nu -1)\approx 0.2$. That in the semidilute regime both
off-lattice and BFM results coincide reinforces our
belief that the mapping between the models is robust. The divergence of the
lattice and off-lattice data at very high densities is not problematic. A
discussion of this issue we postpone to Sec.~\ref{subsec:LinMelt}.

Within the semidilute regime, {\em i.e.}, within the validity of 
eq.~(\ref{eq:fLsemi}), the mean degree of polymerisation of the linear chains is
power law function of $\phi $ and $\mbox{$\phi_1$}$, and may be brought
under the generic form of eq.~(\ref{eq:NLgeneric}) 
\begin{equation}
N_{s}\equiv \left\langle N_{1}\right\rangle =A_{s}\mbox{$\phi_1$}l^{3\eta
}\phi ^{\eta -1}\exp (\delta E)
\label{eq:NLsemidilute}
\end{equation}
with $\delta =1/2$, $\eta =(1+(\gamma -1)/(3\nu -1))/2\approx 0.6$, and an
amplitude 
\begin{equation}
A_{s}=\sqrt{l_{p}^{3(\gamma -1)/(3\nu -1)}\exp (f_{s})}\approx 1.1
\label{eq:As}
\end{equation}
with $l_{p}\approx 1.1$. The (very weak) $l_{p}$-dependence arises because 
\mbox{$f_1$} \ depends on $b^{3}\phi $ rather than on $l^{3}\phi $. Note
that in the limit of linear chain dominance ($\mbox{$\phi_1$}/\phi \approx 1$
) one recovers the density dependence of the Restricted Model, {\em i.e}., 
$\mbox{$N_1$}\propto \phi ^{0.6}$ \cite{CC90,WMC98}. We verified the validity
of eqs.~(\ref{eq:NLsemidilute}) and (\ref{eq:As}) by comparison with the
computer simulation data, but do not pause here to elaborate on the details.
In Sec.~\ref{subsec:Linscal} we present the full crossover scaling of 
$
\left\langle N_{1}\right\rangle ,$ covering the entire range of
concentrations from the dilute to the melt regime.

\subsection{Concentrated solutions.}\label{subsec:LinMelt}

As we have discussed in a previous paper on the equilibrium polymerisation
in the absence of ring formation \cite{MWL00}, eq.~(\ref{eq:NLsemidilute})
only holds within the semidilute regime. This implies that there must be a
large number of monomers $g$ within a blob, for otherwise the blob concept
becomes meaningless. Within our off-lattice Monte Carlo approach, we have
probed such high densities that the semidilute description does break down
-- this is the melt regime already alluded to. It appears that we enter the
melt regime if $\phi \geq 0.5$, where a different physical behaviour intervenes,
essentially due to fluid-like correlations resulting from local packing
constraints. Clearly, effects of this nature cannot be expected to arise in
any lattice-based bond fluctuation technique, because of the
presence of an underlying lattice structure suppresses to a great extent
fluid-like correlations.

Our off-lattice Monte Carlo simulations point at a linear relationship
between $\mbox{$f_1$}$ and the concentration material upon entering the melt
regime: $\mbox{$f_1$}(\phi )=B_{0}+B_{1}\phi b^{3}$ with $B_{0}\approx -0.62$
and $B_{1}\approx 1.67$. 
See Fig.~\ref{fig:N1phi}, and in particular the inset.
This also clearly shows that within the lattice model, 
the semidilute regime extends deeply
into the melt regime, that is, even for concentrations where the blobs are
so small as to contain no more than $g\approx 10$ monomers. We attribute
this to the lack of a true fluid structure in the lattice-based model.

We conclude from the above that while eq.~(\ref{eq:fLsemi}) breaks down in
the melt, eq.~(\ref{eq:NLSOL}) still holds (see Sec.~\ref{subsec:Linscal}
below). $\left\langle N_{1}\right\rangle $ and $\phi $ are then no longer
related via a pure power law, but in addition via an exponential enhancement
term: $\left\langle N_{1}\right\rangle \propto \phi ^{1/2}\exp \left[
B_{1}\phi b^{3}\right] $. Direct measurement of $\left\langle
N_{1}\right\rangle $ in our simulations confirms this once more. Indications
for the deviation from the usual power-law behaviour at very high densities
have in fact also been found in earlier simulation studies \cite{Yannick,Kroeger}.
The precise reason for the emergence of the exponential
correction is at this point difficult to give. Theoretically, exponential
corrections of the sort found here have been predicted theoretically for
rigid and semiflexible micelles, but only within a second virial theory
valid at low densities \cite{SC94}. Within the second virial theory, the
exponential density dependence of $\left\langle N_{1}\right\rangle $ arises
from excluded volume interactions between the free ends and the central
parts of the chains. Clearly, although the applicability of the concepts
advanced in \cite{SC94} to our high density system of {\em flexible}
bead-spring chains is dubious, one may surmise that differences in the
packing of the central beads and those near the ends of the chains could
well lie at the root of problem. Further study is definitely warranted.

\subsection{Crossover scaling of the mean length of the linear chains.}
\label{subsec:Linscal}

\begin{figure}[tbp]
\centerline{\epsfig{file=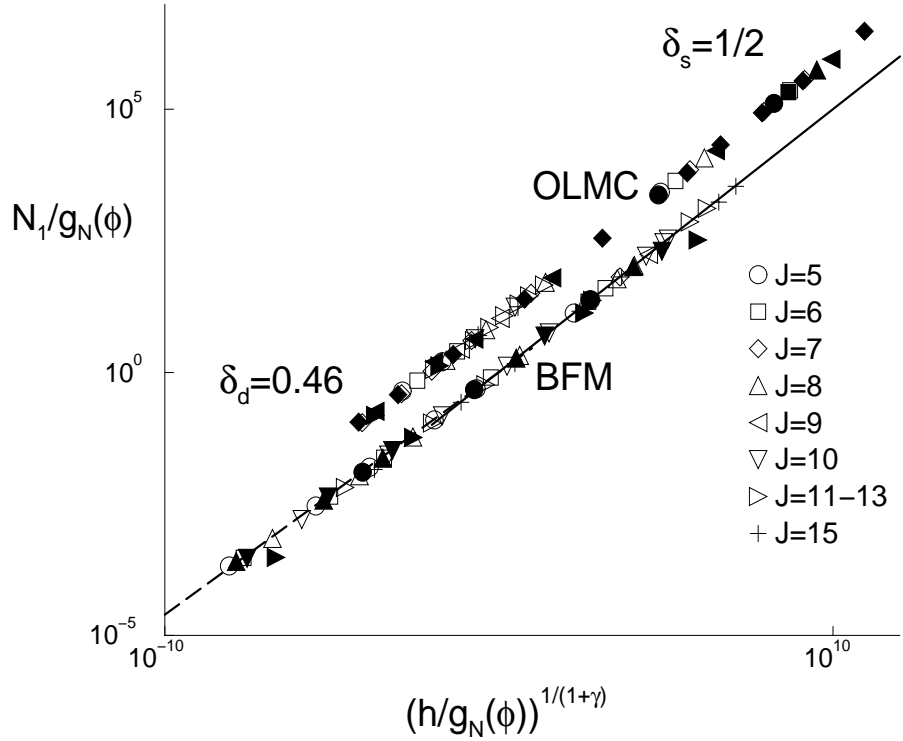,width=120mm,height=100mm}}
\caption{
Scaling attempt for the average aggregation number of linear chains
\NL, to extract the $\delta$
exponent discussed in the main text. The measured values for \NL \
are compared with $g_N(\phi)$ (defined in the text), 
and plotted versus $(h/g_N(\phi))^{1/(1+\gamma)}$, where
$h$ is the (hypothetical) mean chain length for swollen
equilibrium polymers
without {\em inter}-actions between monomers of different chains. The linear
chain monomer density \phiL \ was measured explicitly.
Restricted and Unrestricted Model simulations based on both 
OLMC and BFM methods are shown.
The OLMC data are arbitrarily
shifted upwards for reasons of clarity. The unshifted points all collapse on
the same
master curve! This confirms validity of the mapping between the different
simulation models, and
the universality of the scaling with the quantity $g_N(\phi)$, which we
identified as the
generalized number of monomers per blob containing the directly measured
$\fL(\phi)$ (see Fig.\ref{fig:fLN}). This relation is {\em not}
altered by the presence of the rings. Hence, the free energy of a chain end
is again shown to be a function of $\phi=\phiR+\phiL$ only. 
\label{fig:N1gh}}
\end{figure}

\begin{figure}[tbp]
\centerline{\epsfig{file=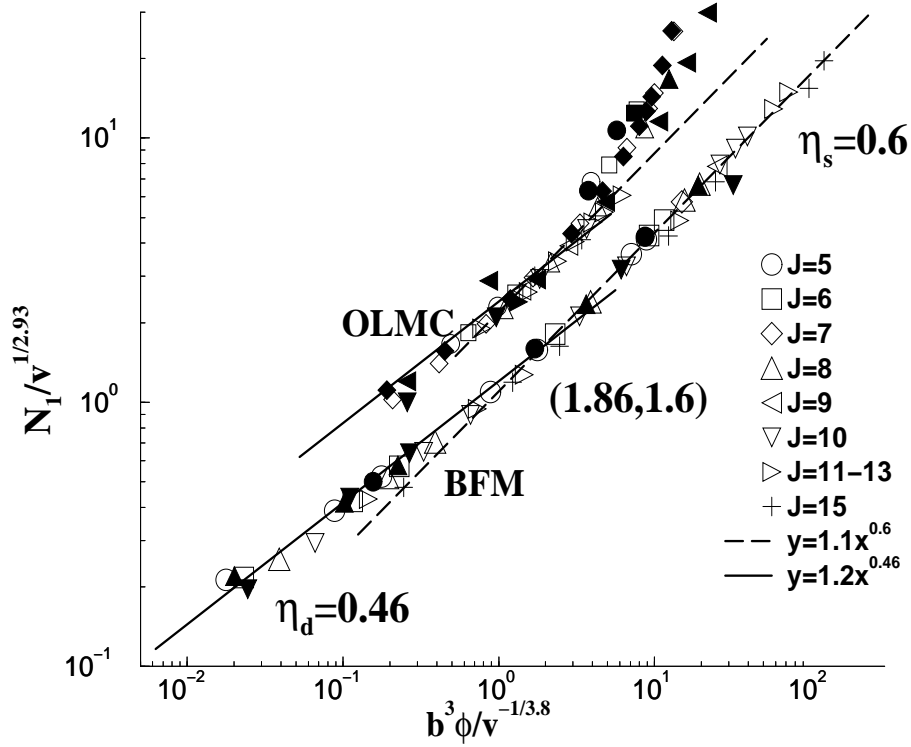,width=120mm,height=100mm}}
\caption{
The scaled mean aggregation number of linear chains $\mbox{$N_1$}$ versus
the scaled density $\phi$, chosen such as to get a scaling with
the $\eta$ exponent. As explained in the text, \mbox{$N_1$} \ and $\phi$ are
rescaled with powers of $v=\exp(E) \mbox{$\phi_1$}/\phi$. In the Restricted
Model (open symbols), and for large densities in the Unrestricted Model,
$v$ reduces to the density-independent affinity $v=\exp(E)$.
Again,  the OLMC data are shifted
arbitrarily upwards for clarity.}
\label{fig:N1v}
\end{figure}

In this subsection we discuss the crossover scaling behaviour for the linear
chain aggregation number $\left\langle N_{1}\right\rangle $, focusing on the 
$\delta $-exponent (Fig.~\ref{fig:N1gh}) and on the $\eta $-exponent 
(Fig.~\ref{fig:N1v}). We show that the computer simulation data from both
Restricted and Unrestricted Models collapse onto the same universal
function, if properly rescaled, and the directly measured density 
$\mbox{$\phi_1$}$ is used. Again we include data from both the bond
fluctuation model (BFM) and the off-lattice Monte Carlo (OLMC) method.
Results from the latter technique have been shifted upwards for reasons of
clarity in both the figures \ref{fig:N1gh} and \ref{fig:N1v}.

In Fig.~\ref{fig:N1gh} we compare the actual, measured $\left\langle
N_{1}\right\rangle $ with a ``hypothetical'' dilute mean chain length 
$h(\mbox{$\phi_1$})$ calculated from eq.~(\ref{eq:NLdilute}). In the dilute
regime $h\equiv N_{d}$, but outside this regime $h$ represents an
extrapolated value. Plotted is not $\left\langle N_{1}\right\rangle $
against $h$, but the quantity $y=\left\langle N_{1}\right\rangle /g_{N}$
against the quantity $x=(g_{N}(\phi )/h(\mbox{$\phi_1$}))^{1/(1+\gamma )}$,
specifically chosen to get a scaling with the $\delta $-exponent: 
$y\propto x^{\delta }$. Here, $g_{N}$ is the mean chain length at the
crossover density from the dilute to the strongly overlapping regime,
defined by the equality $g_{N}\equiv N_{d}=N_{SOL}$. 
Equating eqs.~(\ref{eq:NLdilute}) and (\ref{eq:NLSOL}), gives for this quantity 
\begin{equation}
g_{N}(\phi )=A_{d}^{1+2/(\gamma -1)}\exp (-\mbox{$f_1$}(\phi )/(\gamma -1))
\label{eq:gNdef}
\end{equation}
In the semidilute regime $g_{N}$ becomes proportional to $g$, with an
amplitude $G_{N}=A_{d}^{1+2/(\gamma -1)}\exp \left[ -f_{s}/(\gamma
-1)\right] \approx 5$ very close to the blob prefactor $G=5.2$. Because of
the very large exponent $2/(\gamma -1)\approx 12.7$, we regard these two
values as numerically identical. Our conclusion is that $g_{N}(\phi )$
represents a generalization of the number of monomers per blob $g(\phi )$,
also valid in the limit of very high densities where eq.~(\ref{eq:fLsemi})
does no longer apply. In passing we also note that $G_{N}$ is also
consistent with $G_{1}$ ({\em cf}. eqs.~(\ref{eq:As}) and (\ref{eq:G1})). In
other words, our characterisation of the excluded-volume blob is internally
consistent.

The purpose of all this is to make clear that in the high density regime
discussed in the previous section, the mean aggregation number $\left\langle
N_{1}\right\rangle $ can be described by the same universal function that is
valid in the dilute and semidilute regimes, if the notion of the number of
monomers per blob is generalised appropriately. The function $g_{N}$ that
describes the generalised blob is universal in the sense that it depends
solely on the overall monomer density $\phi$, and on the number of monomers
per effective step length $l_{p}$. 
Indeed, the data points obtained for the Restricted
and Unrestricted Models, within the bond fluctuation and off-lattice Monte
Carlo methods, all collapse on the {\em same} universal function. The
scaling plot confirms the scaling relation we set out to investigate 
$y\propto x^{\delta }$, with $\delta =0.46$ in the dilute limit, and 
$\delta =1/2$ in the semidilute and melt limits.


In our second scaling plot, Fig.~\ref{fig:N1v}, we show the effects of a
density variation on $\left\langle N_{1}\right\rangle $. To focus on the
concentration behaviour of the critical exponent $\eta $, we make use of
eqs.~(\ref{eq:NLdilute}) and (\ref{eq:NLsemidilute}), and plot 
$y=\left\langle N_{1}\right\rangle /\upsilon ^{1/\kappa }$ versus $x=\phi
b^{3}\upsilon ^{1/\varphi }$ where $\upsilon \equiv \exp \left( E\right)
\phi _{1}/\phi $, and $\kappa =(3\nu -1)\varphi \approx 2.93$ and $\varphi
=1+(1+\gamma )/(3\nu -1)\approx 3.81$ are known critical exponents. This
should yield a scaling relation $y\propto x^{\eta }$. Equating equations 
(\ref{eq:NLdilute}) and (\ref{eq:NLsemidilute}) gives a crossover density 
$l^{3}\mbox{$\phi^{*}$}=Pv^{-1/\varphi }$ and a crossover length 
$\mbox{$N^{*}$}=Qv^{1/\kappa }$. The amplitudes $P=(A_{d}/A_{s})^{1/(\eta
_{s}-\eta _{d})}\approx 1.9$ and $Q=A_{s}P^{\eta _{s}}\approx 1.6$ are
determined from $A_{d}$ and $A_{s}$, exactly as in the case of the
Restricted Model, where $\upsilon =\exp (E)$\cite{WMC98}. The estimates
cannot be expected to be very accurate because of the large exponent 
$1/(\eta _{s}-\eta _{d})=7.14$, but are consistent with the estimates
obtained by using the amplitudes $A_{d}$ and $A_{s}$ predicted from $G_{1}$
(and/or $G$.)

The data points for the dilute and semidilute regimes clearly collapse. As
discussed earlier in the context of Fig.\ref{fig:N1phi}, the off-lattice
Monte Carlo data do not conform to the scaling theory at very high
densities, due to the effects of packing that dominate the melt regime. 

\section{Mass distributions of closed loops.}\label{sec:Ring}

It has become clear from the discussion of Sec.~\ref{sec:Size}, that the
dimensions of the linear equilibrium polymers are successfully described in
terms of the Flory exponent $\nu $. In the preceding Sec.~\ref{sec:Lin} we
demonstrated that another critical exponent, the susceptibility exponent 
$\gamma $, is the exponent relevant to the description of the length
distribution of the linear chains in the dilute and semidilute regimes. The
length distribution of rings, only formed in the Unrestricted Model, is
dominated by yet another critical exponent, $\alpha $. This exponent is the
well-known specific heat exponent, related to the Flory exponent $\nu $ 
{\em via} the hyperscaling relation $\alpha =2-D\nu $ in $D$ spatial dimensions.
As we shall make plausible below, the ring distribution is quite accurately
described by the quasi singular scaling relation $\mbox{$c_0$}(N)\propto
N^{-(3-\alpha )}$.

\begin{figure}[tbp]
\centerline{\epsfig{file=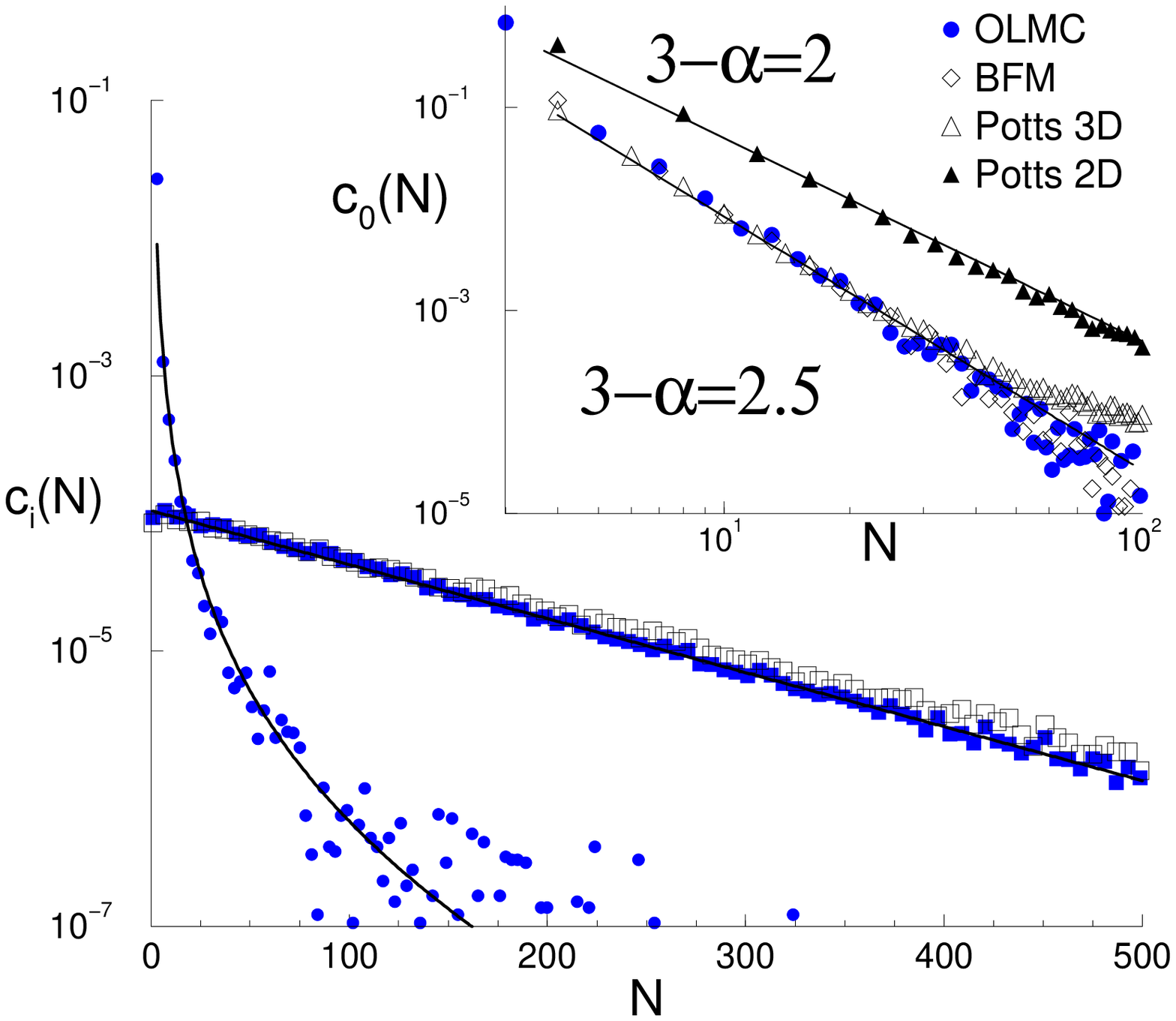,width=120mm,height=100mm}}
\caption{
Probability Distributions Functions of self-assembled chains in the Strong
Overlap Limit (SOL). Main figure: Comparison of the computer simulation
results for the Restricted Model (RM)
and  the Unrestricted Model (UM), obtained by the off-lattice Monte Carlo
method (OLMC), described in the main text. The open squares represented the
distribution of linear chains in the RM, the filled symbols
the distribution of rings (circles) and that of linear chains (squares)
within the UM. 
Data shown are from simulations at a monomer density of $\phi=1.5$, 
and scission energy of $J=7$.
At such a high density, the UM system is dominated by linear chains
($\mbox{$\phi_1$}/\phi\approx 0.88$), and the distribtutions $\cL(N)$ of
both models are virtually identical exponentials
$\mbox{$c_1$}(N)\propto \exp (-N/\mbox{$N_1$})$, where for convenience the
notation $\mbox{$N_1$}\equiv \langle \mbox{$N_1$} \rangle$ for the mean
aggregation number of linear chains is used. For our choice of density and
scission energy,
($\mbox{$N_1$}^{RM}\approx 120$, $\mbox{$N_1$}^{UM}\approx 111$).
The ring distribution is well described by the power law
$\mbox{$c_0$}(N) \propto N^{-(3-\alpha)}$, with a slope $3-\alpha=2.5$ (bold
line). 
Inset: $\mbox{$c_0$}(N)$ in a double logarithmic plot for the
OLMC, BFM and Potts Model algorithm (vertical axes shifted for clarity).
}
\label{fig:cLRN}
\end{figure}

Let us return to main Fig.~\ref{fig:cLRN}. The distribution of rings indeed
follows a power law distribution with an $\alpha =0.5$ in the dense limit of
the off-lattice Monte Carlo simulation. This we expect, because in the dense
limit $\nu =1/2$ and the simulation was done in $D=3$. The hyperscaling
relation is checked extensively in the inset of the figure, where we plot
the distributions as obtained for $D=3$ not only by means of the off-lattice
and bond fluctuation methods, but also from the grand-canonical Potts model
simulations, mentioned in Sec.~\ref{sec:Algo}. Also included are data from
the Potts model in two dimensions, where $\alpha =1$. Because the Flory
exponent is $\nu =1/2$ for the concentrated or melt regime in both two and
three dimensions, our results confirm the hyperscaling relation.

\begin{figure}[tbp]
\centerline{\epsfig{file=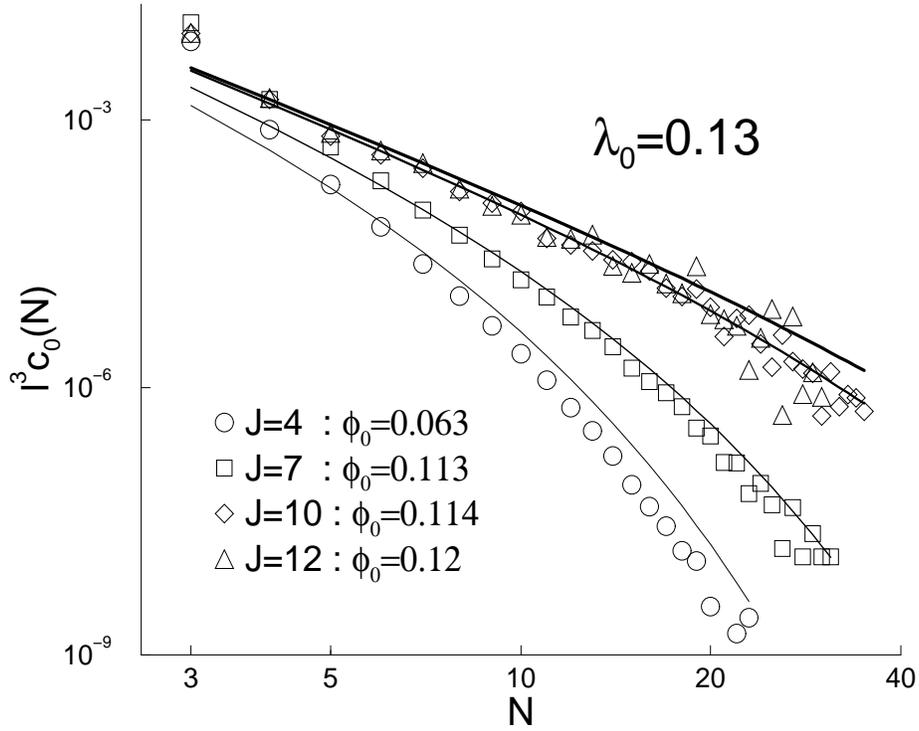,width=120mm,height=100mm}}
\caption{
Ring distribution (symbols) obtained from the off-lattice Monte Carlo method
for the Unrestricted Model in the dilute limit at a single concentration
$\phi=0.125$, but for different values of the scission energy $J$.
Double-logarithmic plot. The fits to the theoretical curves discussed in the
text (and indicated by drawn lines), confirm that the ring distribution
$\mbox{$c_0$}(N)$ depends on $E$ only indirectly via $\mu$, i.e., via
\mbox{$N_1$}.  The plots confirm the hyperscaling relation $\alpha+D\nu=2$,
and fix the prefactor of the scaling theory
$\lambda_0 \approx 0.1$ in the dilute limit. }
\label{fig:c0phi0.125}
\end{figure}

The strong power law behavior seen in Fig.~\ref{fig:cLRN} does not exclude
an additional exponential damping, which we expect to be present from 
eq.~(\ref{eq:cLRN}). In the strong overlap regime this exponential damping is
difficult to detect, because there $\mu =1/\left\langle N_{1}\right\rangle $
is very small. However, the exponential damping of the ring distribution is
observable in the dilute regime, as is made clear by Fig.~\ref
{fig:c0phi0.125}. In the figure we present data from the off-lattice
simulations, for different scission energies $J$ and at a fixed
concentration $\phi =0.125$. Note that $\mbox{$\phi_0$}$ increases with $J$,
as does $\mbox{$N_1$}=\gamma /\mu $ (see Sec.~\ref{subsec:LinDilute}). For
large values of $J$ we recover the power law behavior seen in 
Fig.~\ref{fig:cLRN}, albeit with a different exponent $\alpha =3-2.764$ because
interactions are now not screened. This value is again in line with the
hyperscaling relation, since in the dilute regime the Flory exponent obeys 
$\nu =0.588$. The drawn lines are fits eq.~(\ref{eq:cRN}).

\begin{figure}[tbp]
\centerline{\epsfig{file=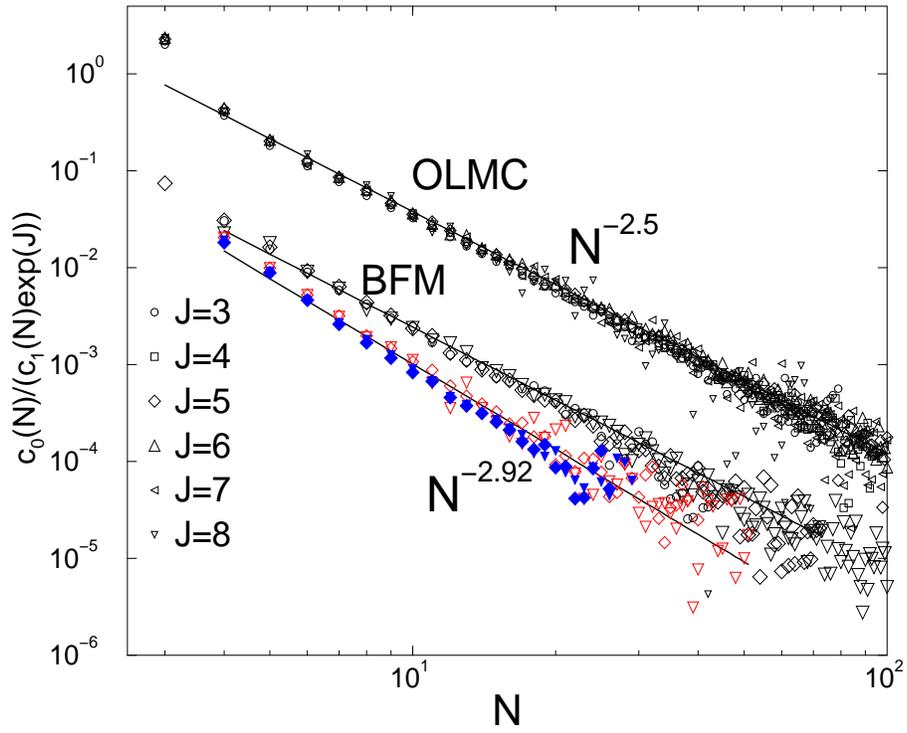,width=120mm,height=100mm}}
\caption{
Power-law behavior of the ratio of the length distributions for rings and
linear chains, $\mbox{$c_0$}/\mbox{$c_1$}$,
versus the aggregation number $N$ (OLMC vertically shifted).
The slopes conform to a power law exponent
$\tau=2.5$ in the strong overlap limit, and a $\tau=2.92$ in the dilute
limit. }
\label{fig:Z0Z1}
\end{figure}

If the properties of rings and linear chains really decouple as was assumed
in our Flory-Huggins ansatz, we can measure the free energy difference
between rings and linear chains directly by plotting the ratio 
\begin{equation}
\frac{\mbox{$c_1$}(N)}{\mbox{$c_0$}(N)}=\exp (\mbox{$f_0$}(N,\phi )-(E+
\mbox{$f_1$}(N,\phi ))).
\label{eq:ratiopart}
\end{equation}
versus the aggregation number $N$. 
This is done in Fig.~\ref{fig:Z0Z1}, where
we present off-lattice Monte Carlo data taken at a density of $\phi =1.5$,
and BFM data at $\phi =0.0625$ (middle curve), the former
shifted vertically for reasons of clarity. Both densities are in the strong
overlap regime. The lowest curve gives BFM data, taken at
the dilute densities $\phi =0.0125$ (open symbols) and $\phi =0.00125$
(filled symbols). The lines indicate power laws 
$\exp (\mbox{$f_1$}- \mbox{$f_0$})\propto N^{-\tau }$. 
From what we have said above about $\alpha 
$, we expect $\tau =\gamma +D\nu =2.92$ in the dilute limit, and $\tau
=1+D/2=2.5$ in the strong-overlap limit. The Monte Carlo data confirm this
expectation, demonstrating once more the validity of the Flory-Huggins
ansatz (task 2).

\begin{figure}[tbp]
\centerline{\epsfig{file=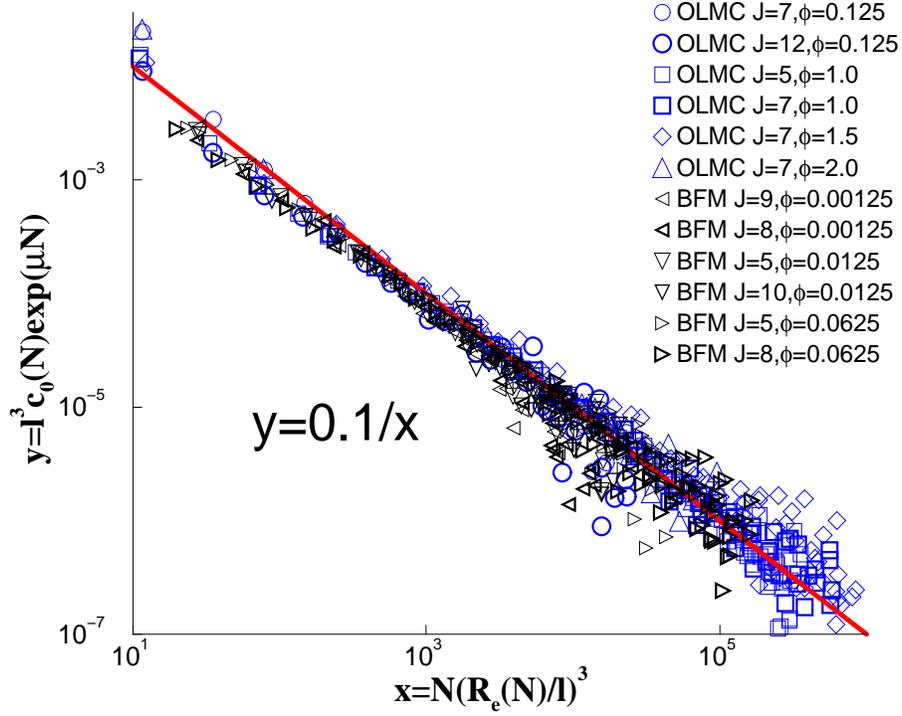,width=120mm,height=100mm}}
 \caption{
Scaling plot of the quantity $y=l^3\mbox{$c_0$}(N)\exp(\mu N)$ versus
$x=N (R_{e1}(N)/l)^3)$, using the directly measured end-to-end distrance
$R_{e1}(N)$ of the linear
chains for various systems for different algorithms and regimes, as
indicated
in the figure. The data collapse onto a master curve $y=\lambda_0/x$ with
$\lambda_0\approx 0.1$.}
\label{fig:c0scal}
\end{figure}

Let us now briefly pause at a simple argument due to Porte, by which one may
derive an expression for the ring distribution, and arrive at the exponent 
$\alpha $ \cite{Porte}. The ratio of the ring and linear chain distribution
functions, eq.~(\ref{eq:ratiopart}), must be equal to the ratio of the
respective partition functions, which in turn must proportional to the
probability of opening a loop. The probability of opening a loop is
proportional to (i) the Boltzmann weight $\exp (-E-\mbox{$f_1$})$ to break a
single bond\cite{com:bothsides}, (ii) the number of places where the ring
can break, $N$, and (iii) the volume $R_{e1}^{3}$ that two neighboring \
segments can explore after being disconnected. Hence, 
\begin{equation}
l^{3}\mbox{$c_0$}(N)=\lambda _{0}\frac{\exp (-\mu N)}{N\left( R_{e1}(N,\phi
,l_{p})l^{-1}\right) ^{3}}H(N-\mbox{$N_c$}).
\label{eq:cRN}
\end{equation}
with $\lambda _{0}$ an unknown constant of proportionality. We have put 
eq.~(\ref{eq:cRN}) to the test in Fig.~\ref{fig:c0scal}, using the directly
measured end-to-end distance $R_{e1}(N,\phi )$ discussed in Sec.~\ref
{sec:Size}, and the measured chemical potential $\mu $, as discussed in 
Sec.~\ref{sec:Lin}. A wealth of data from both the lattice-based and off-lattice
Monte Carlo methods in various regimes is included in the figure. The
collapse of the data is next to perfect -- the main result of this paper.
The non-trivial behaviour of $R_{e1}(N,\phi )$ in the dilute and semidilute
regimes explains the complex density dependence of the ring distribution in
the various concentration regimes. The scaling plot yields a value of 
$\lambda _{0}\approx 0.1$ for the constant of proportionality 
(Fig.~\ref{fig:c0phi0.125}), 
similar to the one estimated by Pfeuty and co-workers\cite{Petschek} in their 
analysis of self-assembled chains.

\begin{figure}[tbp]
\centerline{\epsfig{file=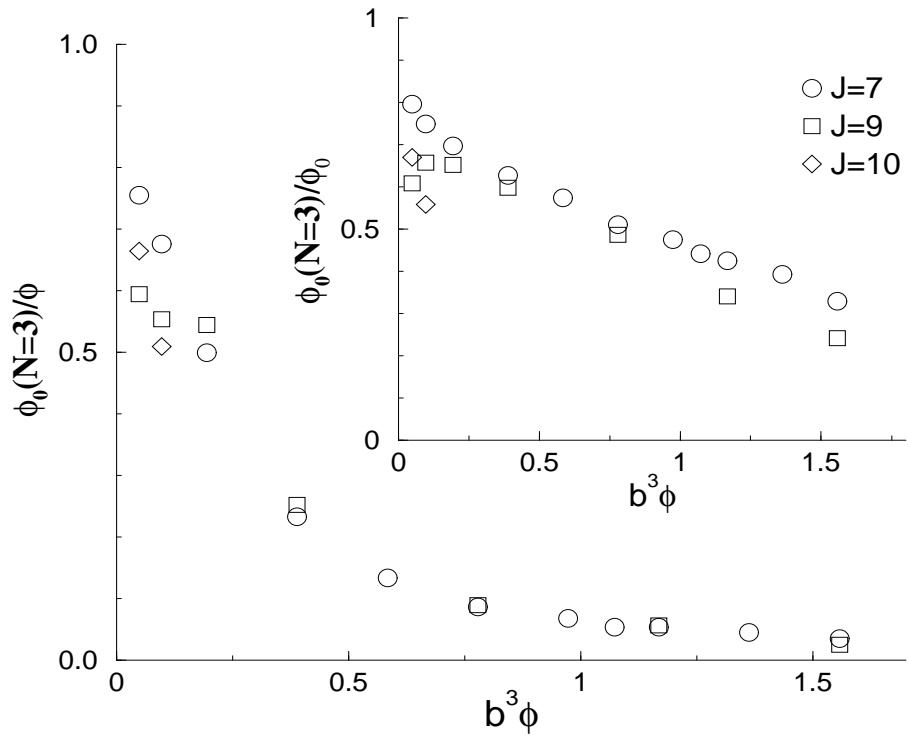,width=120mm,height=100mm}}
\caption{
Variation of the fraction material in trimers
$\mbox{$\phi_0$}(N=3)/\phi$ with overall monomer density $\phi$, as found in
the OLMC simulations. Inset: variation of the fraction monomers in trimer
rings, relative to that in all rings \mbox{$\phi_0$}, versus the overall
density of monomers. For all but the highest density, the
mass is mainly concentrated in tiny rings. The $J$-variation is very weak. }
\label{fig:phi3phi}
\end{figure}

While the scaling relation eq.~(\ref{eq:cRN}) appears to be well satisfied
for relatively large $N$, strong deviations from the universal asymptotic
behavior is observed for small $N$. As seen from inspection of, e.g., 
Fig.~\ref{fig:c0phi0.125} or Fig.~\ref{fig:Z0Z1}, an unexpectedly large number of
trimer rings are present in our simulations. 
This is emphasized in Fig.~\ref{fig:phi3phi}, 
where we give the fraction of monomers trimer rings 
$\mbox{$\phi_0$}(N=3)$ relative to the total amount, and to that in all the
rings, as function of the overall concentration of aggregating material. At
densities below $\phi =0.25$ most of all monomers are contained in trimers!
The fraction of trimer rings seems largely independent of $J$.

\begin{figure}[tbp]
\centerline{\epsfig{file=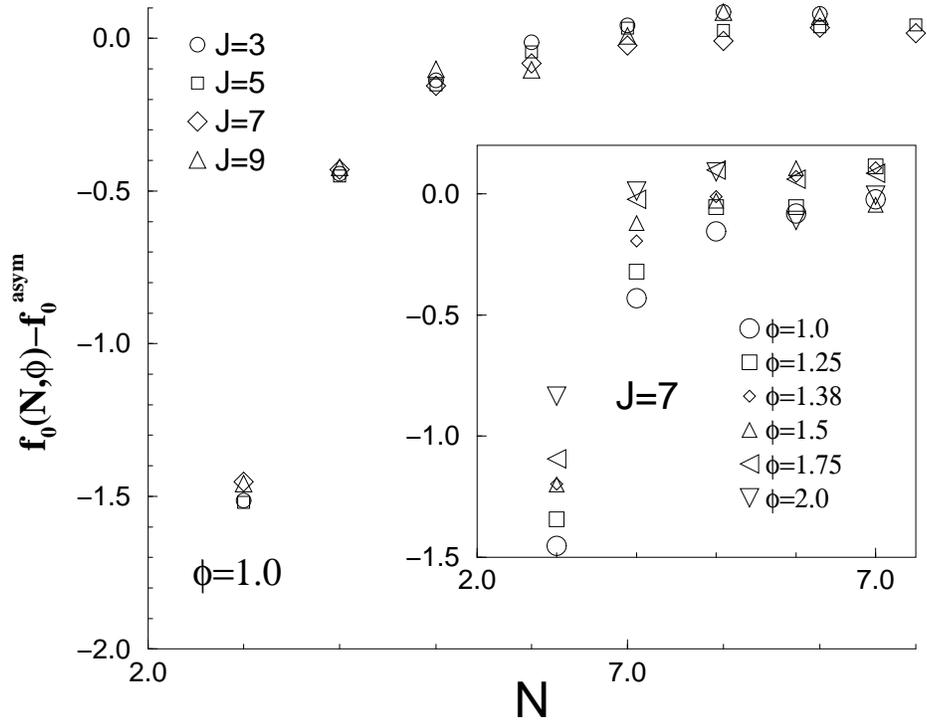,width=120mm,height=100mm}}
\caption{
Small-ring effect for in the Unrestricted Model. Data were taken by means of
OLMC simulations, in the strong overlap limit. Shown is the free energy
difference
$\mbox{$f_0$}(N,\phi)-\mbox{$f_0$}^{asym}$ versus the aggregation number
$N$, where $\mbox{$f_0$}=-\log(\mbox{$c_0$})-\mu N$ and
$\mbox{$f_0$}^{asym}$
represents the known asymptotic value for large $N$. Main figure: Results
for the density
$\phi=1.0$, and various scission energies $J$. Inset: Similar to main
figure, but now a fixed scission energy $J=7$ and various densities
$\phi$. }
\label{fig:fRN}
\end{figure}

The effects of small chain length are systematically analyzed in 
Fig.~(\ref{fig:fRN}), 
where we compare the measured free energy $\mbox{$f_0$}(N,\phi)$
with the asymptotic behavior $\mbox{$f_0$}^{aym}=a-(1+D\nu )\log (N)$, valid
for large $N$ . Here, $a$ is a constant function of $N$, although it does
depend on $\phi$, because of the density dependence $R_{e1}$ in the SOL. 
The density dependence shown in the inset, points at a decreasing tendency 
to form small rings with increasing concentration $\phi $. 
It appears that the effect is much stronger in the
off-lattice simulations than in the lattice-based approach, which reaches
the asymptotic limit more rapidly (results not shown).

It is important to emphasize that the observed small-ring behaviour is not
in contradiction with the scaling picture presented in Fig.~\ref{fig:c0scal}. 
It is in part caused by the similar deviation from the asymptotic
behaviour of the dimensions of short linear chains, as can be seen in 
Fig.~\ref{fig:RJ8N}. We find that short chains are smaller in dimension than
expected from the scaling behaviour for large $N$. As the entropy gain of
opening a ring is smaller the shorter the chain, short rings become as a
result of this more likely than expected. (See eq.~(\ref{eq:cRN}).) In other
words, the deviation from the asymptotic behavior of $\mbox{$c_0$}(N)$ at
small $N$ is caused (at least in part) by the small-$N$ behaviour of the
dimensions of the linear chains.

That there are deviations from the universal asymptotic behavior in the
small-ring population does not really come as a surprise. Unfortunately,
even seemingly small shifts in the distribution of rings do matter when it
comes to determining the relative amounts of rings and linear chains: the
essentially algebraic distribution forces most of the ring mass to be
concentrated in the smallest rings --- exactly where the universal
description based on scaling laws becomes inaccurate. This is the reason why
obtaining a universal diagram of states based on theoretical argument,
attempted in the next section, is fraught with difficulty, at least in
principle. However, for the purpose of getting a qualitative picture of the
aggregated states of equilibrium polymers, scaling theory provides a
sufficiently accurate basis.

\section{Diagram of states.}\label{sec:PD}

If we accept the theoretical distributions of eq.~(\ref{eq:cLRN}) at face
value, and augment these with the values for the parameters $b$, $G$, $G_{1}$, 
$A_{d}$, $A_{s}$ and $\lambda _{0}$ obtained by fitting to the results of
our computer simulations, we can calculate a diagram of states or ``phase
diagram'' for our system of flexible equilibrium polymers \cite{com:PD}. For
this purpose, we calculate, using eq.~(\ref{eq:cLRN}), 
the overall densities
of monomers in rings and in linear chains, $\mbox{$\phi_0$}$ and 
$\mbox{$\phi_1$}$, as well as the mean chain lengths, $\left\langle
N_{0}\right\rangle $ \ and $\left\langle N_{1}\right\rangle $,\ as a
function of the total monomer density $\phi $, and of the end cap energy $E$. 
For the mean end-to-end distance \mbox{$R_e$},\ and the free energy
correction \mbox{$f_1$}\ to create an additional chain end, we simply put 
\begin{eqnarray}
\mbox{$R_e$}(N,b^{3}\phi ) &=&\min (bN^{\nu },bg^{\nu }(N/g)^{1/2}) \\
\label{eq:approx}
\mbox{$f_1$}(N,b^{3}\phi ) &=&\max (-(\gamma -1)\log (N),-(\gamma -1)\log
(g_{1}))\nonumber 
\end{eqnarray}
with the monomer number per blob $g(b^{3}\phi )=5.2(b^{3}\phi )^{-1/(3\nu
-1)}$ and $g_{1}(b^{3}\phi )=0.083g(b^{3}\phi )$. Obviously, 
eq.~(\ref{eq:approx}) is only an approximation to the full universal functions.

The set of equations to be solved require a numerical evaluation,
essentially because the sums cannot all be evaluated analytically. For a
given $E$ we vary the Lagrange multiplier $\mu $, and obtain from 
eq.(\ref{eq:cLRN}) the densities \mbox{$\phi_0$}, \mbox{$\phi_1$}\ and, therefore,
also $\phi $. A complication is that the distribution functions themselves
depend explicitly on the density $\phi $, at least in the semidilute regime.
The total density is of course not known {\em a priori}, and has to be
evaluated for any given value of $\mu $. In principle, the only way out is a
recursive iterative scheme. Fortunately, matters become simplified
significantly if one starts the calculation in the dilute regime (at large 
$\mu $) where both \mbox{$R_e$}\ and \mbox{$f_1$}\ are density independent.
Recursive iteration may then be circumvented by slowly decreasing the value
of $\mu $, updating $\phi $ and $g(\phi )$, and then using a forward scheme
to determine the next $\mu $-value.

\begin{figure}[tbp]
\centerline{\epsfig{file=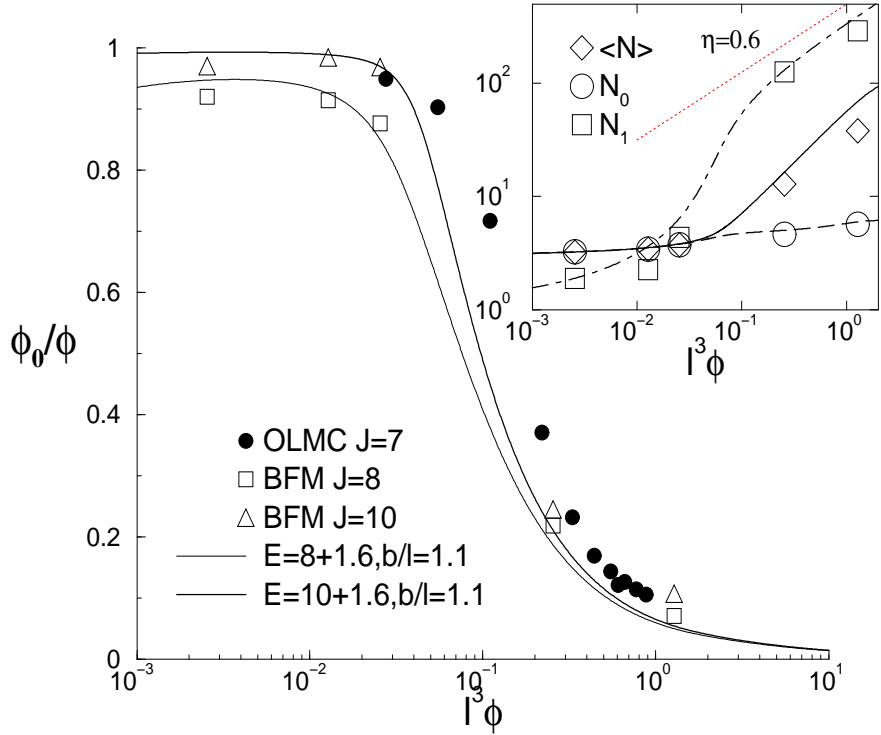,width=120mm,height=100mm}}
\caption{
Comparison of the concentration dependence of the fraction of rings as
obtained by Monte Carlo simulations, and that calculated from the theory,
described in the main text. The curves
(for $\lambda_0=0.1$, $\delta J=1.6$ and $l_p=1.1$) contain no additional
fit
parameters! Main figure: Ratio $\mbox{$\phi_0$}/\phi$ of monomers contained
in rings. Data from BFM (open symbols) and OLMC simulations
(filled symbols) are indicated.
Inset: $\mbox{$\langle N \rangle$}$, \mbox{$N_0$}\ and \mbox{$N_1$}\ for
$J=10$.
Obviously, $\mbox{$\langle N \rangle$} \approx \mbox{$N_0$}$ for
$\mbox{$\phi_0$}/\phi\approx 1$, and for large $\phi$ we have $\mbox{$N_1$}
\propto
\phi^{-\eta}$ with $\eta = 0.6$.
Note that the growth of \mbox{$\langle N \rangle$} \
is intricate, and could in intermediate regimes lead to a power law with
exponent  $\eta > 0.6$, in agreement with
some (but not all) experimental findings \protect\cite{Schurtenberger}. }
\label{fig:phi0phi}
\end{figure}

We first consider the relative distribution of the monomers over the rings
and the linear chains in flexible equilibrium polymers, where we set the
lower ring cut-off at $\mbox{$N_c$}=3$, conform our simulations. 
In Fig.~\ref{fig:phi0phi} we compare the concentration dependence of the fraction of
monomers in rings, $\mbox{$\phi_0$}/\phi $, as obtained from our simulation
studies, with the numerical results of the idealized model defined above. In
the inset we furthermore give the concentration dependence of the mean chain
lengths $\left\langle N\right\rangle $, $\left\langle N_{1}\right\rangle $
and $\left\langle N_{0}\right\rangle $. Considering that no additional fit
parameters were used, the agreement, although not perfect, is actually quite
reasonable. The general trend is well described by the model calculation.

\begin{figure}[tbp]
\centerline{\epsfig{file=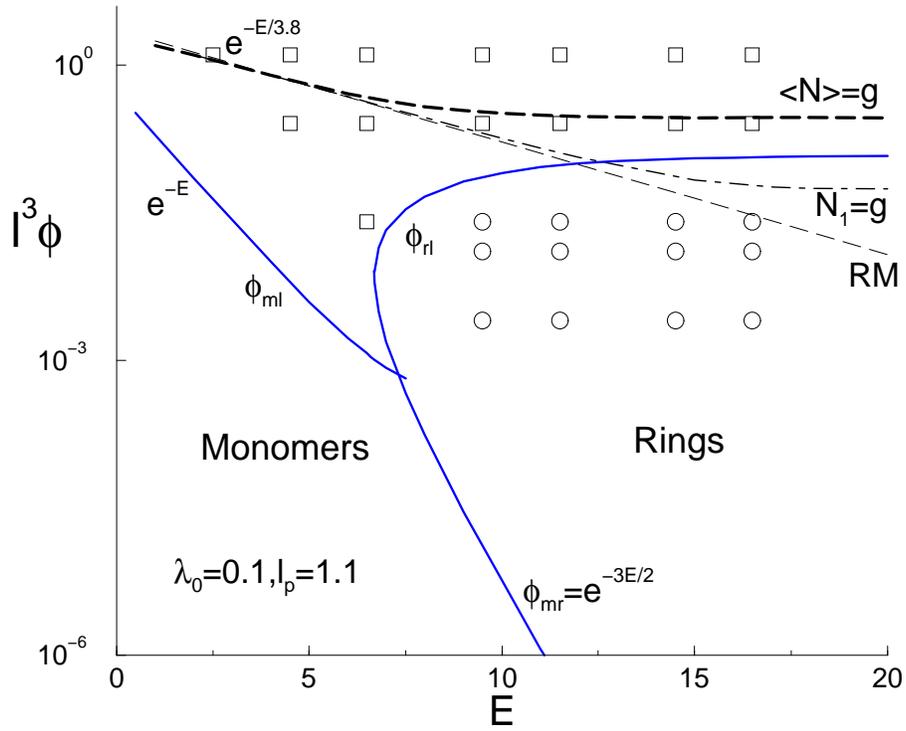,width=120mm,height=100mm}}
\caption{
A diagram of states for the Unrestricted Model, with $\lambda_0=0.1$ and
$l_p=1.1$.
Indicated are three regimes where monomers, rings and linear chains
dominate the population of aggregates, as explained in the text. The ring
regime does not extend to within
the semidilute regime above the $\langle N \rangle = g(\phi)$-line. Also
shown are for what values of $E$ and $l^3\phi$ we find in the BFM
simulations configurations where $
\mbox{$\phi_0$}/\phi > 1$ (spheres), and those where $\mbox{$\phi_1$}/\phi <
1$
(squares). }
\label{fig:pdEV}
\end{figure}

The full diagram of states is presented in Fig.~\ref{fig:pdEV}. We
distinguish three regimes in the $(E,\phi )$-plane: (i) one where free
monomers dominate at low densities and end-cap energies, (ii) one where
rings dominate, at intermediate densities and high end-cap energies, and
(iii) one where linear chain dominate, at high densities. The crossover to
the ring-dominated regime is defined by the equality 
$\mbox{$\phi_0$}=\mbox{$\phi_1$}=\phi /2$. 
For energies larger than $E=6.7$, we find {\em two} roots: 
the concentration $\phi _{rl}$ demarcating the crossover between ring
and linear chain dominance, and the concentration $\phi _{mr}$ separating
the monomer- and ring-dominated regimes. The crossover concentration $\phi
_{ml}$ is defined as that concentration when half the linear chains is in
monomeric form, and half in polymeric form of $N>1$. For comparison we have
also indicated in Fig.~\ref{fig:pdEV} data found by means of computer
simulation within the BFM. Configurations with ring
dominance are denoted by circles, systems where linear chains dominate by
squares. Apparently, the simulation data corroborate our theoretical phase
diagram.

In the figure \ref{fig:pdEV} we have in addition drawn the three
dilute-semidilute crossover densities $\mbox{$\phi^{*}$}(E)$ of relevance to
our discussion. The thin dashed line denotes the crossover density 
$l^{3} \mbox{$\phi^{*}$}(E)\simeq 2\exp (-E/3.8)$, 
valid within the Restricted Model (where rings are absent). 
The presence of rings alters qualitatively the crossover density, 
for in the Unrestricted Model it becomes independent
of $E$ at large end-cap energies. This bounds the semidilute regime to
relatively high densities in the Unrestricted Model, as was in fact already
remarked by Cates \cite{CC90}. See Fig.~\ref{fig:pdEV}. The semidilute regime
cannot extend itself deeply into the ring regime, basically because the
rings are too short. This is shown by the dashed-dotted line, which
indicates the crossover density for the linear chains, as defined by
equating $g(\mbox{$\phi_1^{*}$})$ with $\left\langle N_{1}\right\rangle $.
(This definition of \mbox{$\phi_1^{*}$}\ corresponds to the one given in 
Sec.~\ref{sec:Lin}.) Note that $l^{3}\mbox{$\phi_1^{*}$}\rightarrow 0.055$, and
that the chains start to overlap slightly below the $\phi _{rl}$-line. The
rings shorter than $g$ remain swollen at densities well above $\phi _{rl}$.
This can be seen from the thick dashed line, the crossover density
calculated by equating $g(\phi =\mbox{$\phi^{*}$})$ with the mean chain
length $\left\langle N\right\rangle $ of all chains (chains and rings). This
crossover density has an asymptotic value $l^{3}\mbox{$\phi^{*}$}\rightarrow
0.29$ in the limit of large end-cap energies. Around the crossover from the
ring- to the linear chain-dominated regime, the mean chain length grows
rapidly from $\mbox{$\langle N \rangle$}\approx \mbox{$N_0$}\approx 
\mbox{$N_c$}$ to $\mbox{$\langle N \rangle$}\approx \mbox{$N_1$}$, because
then most (but not all) of the additional monomers are now included in the
long chains.

\section{Discussion and Conclusions.}\label{sec:Discussion}

Based on extensive computer simulations in two and three dimensions, using a
number of different simulational techniques, we conclude that the
configurational properties of the self-assembled linear chains are not
altered if ring closure is allowed. The same can be said about the shape of
the probability distribution of the linear chains. We find that the
probability of finding linear chains of a certain length drops off
exponentially with length, at least for concentrations where the chains
strongly overlap. In dilute solution, where the chains do not overlap, the
distribution function is of the Schultz-Zimm form. Previously, the same
distribution functions were found theoretically, as well as by means of
computer simulations, in equilibrium polymerising systems where ring closure
was suppressed. Is ring closure allowed, we find the resulting distribution
of rings to be essentially algebraic, albeit that in dilute solution the
algebraic distribution is dampened an additional exponential length
dependence.

It seems that the rings formed in equilibrium polymerisation merely deplete
monomers from solution, making fewer of them available for absorption into
linear chains. This is experimentally relevant, because it shifts the
crossover to the semidilute regime to significantly higher densities.
Interestingly, of the amount of material present in the rings, a very large
fraction resides in the smallest rings allowed -- in fact much more so than
expected from the universal scaling relations which seem to be valid for
large enough rings. In our case the smallest rings possible are determined
by an arbitrary cut-off. In reality, the smallest possible ring is likely to
be determined by the chemistry of the system in hand, making the cut-off a
non-universal quantity. An important factor determining the minimum size of
a ring may well be the rigidity of the bonds connecting the basic building
blocks of the equilibrium polymers, and possibly also the configurational
properties of the building blocks themselves.

It has been argued in the literature that rings shorter than roughly a
persistence length are highly unlikely to form\cite{Porte,SW99}.
If it takes many monomers to make one persistence length, ring
closure is suppressed in favour of linear chains. This might be the reason
why rings seem to be unimportant in giant micellar systems, because each
individual surfactant molecule does not contribute a great deal to the
length of an aggregate. For giant micelles the number of monomers required
to grow an aggregate the size of a persistence length may be very large
indeed, even when the aggregate is quite flexible. 
The same need not be true in other types
of equilibrium polymerising system, and for these ring formation must be
relevant. Indeed, it is well known that in liquid sulfur rings play an
important role. We speculate that the short S$_{8}$ rings, which seem to
overwhelmingly dominate the ring regime of liquid sulfur, represent the
aforementioned lower cut-off.

The question arises why the problem of ring closure has not provoked a great
deal of interest, for instance in relation to other types of equilibrium
polymeric solutions. A plausible reason is that it appears to be difficult
to experimentally distinguish between rings and linear chains, although new
experimental techniques may change this in the future. Meijer and co-workers
have recently been able to distinguish by means of NMR methods between
material in tight rings and that in other chains in their system of
hydrogen-bonded equilibrium polymers\cite{Folmer}.
Unfortunately, these authors have not yet
undertaken any systematic studies of the ring formation, for the moment
barring a comparison with our computer simulation results.

Direct measurements of entire size distributions are also extremely rare in
the field of equilibrium polymeric systems. In fact, we know of only a
single study that has been published in the literature. Greer and co-workers
quite recently presented size distributions of the living polymerisation of
poly$\left(\alpha \text{-methyl styrene}\right)$ in the solvent
tetrahydrofuran\cite{Greer}. The distributions were obtained by means of size
exclusion chromatography after termination of the polymerisation. The data
very clearly show a distribution that is algebraic for small degrees of
polymerisation, crossing over to an exponential distribution for large
degrees of polymerisation, pointing at the presence of small rings. (The
experiments cannot distinguish been rings and linear chains, so the
distribution contains the sum of contributions from linear chains and rings,
if present.) A mean-field power-law distribution proportional to $N^{-2.5}$
turns out to fit the data quite well at low degrees of polymerisation. 

We are somewhat puzzled by the good agreement with the (mean-field)
prediction, for the equilibrium polymerisation reaction is initiated by
sodium naphthalene. This initiator forms bifunctionally reactive polymeric
anions, which in principle do not allow for a ring closure reaction.
However, there are several mechanisms by which rings may form regardless.
One mechanism is that of Coulombic association of the charged end groups.
This is possible because of the presumably strong binding 
or localisation of the cationic counterions to the anionic end groups, 
the solvent being very hostile to charged species.
The resulting effectively dipolar chain ends may
themselves form larger clusters (dimers, trimers etc.) for the same reason,
allowing in principle for all kinds associations, including rings.
Another mechanism could be the presence of small amounts of oxygen,
which in the termination reaction lead to the formation of polymer radicals
which could close up. Even if rings are indeed formed in the living
polymerisation studied by Greer, we feel a direct comparison with our
simulations is not warranted. The reason is that these rings must then be
formed by processed competing with the on-going linear anionic
polymerisation, requiring a different model altogether.

\section*{Acknowledgments}

This research has been supported by the Bulgarian National Foundation for
Science and Research under Grant No. X-644/1996. JPW thanks M.E.~Cates for
helpful comments and hospitality in Edinburgh.


\newpage

\newpage

\newpage

\newpage

\newpage

\newpage

\newpage

\newpage

\newpage

\newpage

\newpage

\newpage

\newpage

\newpage

\newpage


\begin{references}
\bibitem{com:GM}  
The physical properties of EP and specifically GM
have been reviewed briefly in \cite{WMC98} and more extensively in ref.~\cite
{CC90}. In ref.\cite{WMC98} and \cite{MWL00} we also review and analyze some
of the older computational work on EP.
\bibitem{com:depletion}  
Note that the $\gamma $-exponent
for $D=3$ is only slightly larger than its mean-field vlue $\gamma =1$.
Hence, the predicted power law depletion in the probability distributions
for small $N$ is very weak and requires a relatively large mean mass at low
densities to be seen. This is why within the range of the parameters
accessible essentially only the tail $\mbox{$c_1$}(x)\propto \exp (-\gamma x)
$ is visible.
\bibitem{com:bothsides}  
Note the cancellation of the
Boltzmann weight $\exp (-(E+\mbox{$f_1$}))$ occurring on both sides of this
equation. Hence, $\mbox{$c_0$}$ is independent on both $E$ and 
$\mbox{$f_1$}(N,\phi )$, i.e. on the specific model hamiltonian describing, e.g., the
chain rigidity influencing both terms. This was not appropriately seen by
two of us (PvS,JPW)\cite{SW99}.
\bibitem{com:PD}  
It has to be stressed that ``phase'' means here
``regime'' rather than thermodynamic distinct phases associated with
discontinuities in the free energy of the system. The transitions between
different regimes only foreshadow the true thermodynamic phase transition at
infinite end cap energies.
\end{references}
\end{document}